\documentclass[letterpaper]{article} 

\newif\ifreal\realtrue

\usepackage[T1]{fontenc}
\usepackage[a4paper, margin=1.5in]{geometry}

\usepackage{algorithm}

\newif\iflong\longtrue
\newif\ifwrong\wrongfalse

\newif\iffun\funtrue
\newif\iflncs\lncstrue

\newcommand{\lncsqed}{\iflncs\hfill$\square$\fi}

\usepackage[disable]{todonotes}

\usepackage{cite}
\usepackage{amsmath,amssymb,amsfonts}
\usepackage{graphicx}
\usepackage{booktabs}
\usepackage{url}
\usepackage{amsthm}
\usepackage{comment}
\usepackage{authblk}

\usepackage{cleveref}
\usepackage{xspace}
\usepackage{xcolor,colortbl} 
\usetikzlibrary{calc}
\usepackage{dsfont}
\usepackage{booktabs}
\usepackage{array}

\newcommand{\todog}[2][]{\todo[#1,color=yellow!90!red]{#2}}

\newcommand{\todom}[2][]{\todo[#1,color=green!50]{#2}}
\newcommand{\todomi}[1]{\todom[inline]{#1}}
\newcommand{\todop}[2][]{\todo[#1,color=magenta]{#2}}

\newtheorem{theorem}{Theorem}

\newtheorem{proposition}[theorem]{Proposition}
\newtheorem{definition}{Definition}

\crefformat{observation}{Observation~#1}
\crefformat{myclaim}{Claim~#1}
\newenvironment{claimproof}{{\noindent\textit{Proof. }}}{\hfill$\blacksquare$}
 
\newcommand\job[3][adj]{\mathrm{#1}^{#2}_{#3}}
\newcommand\ajob[2]{\aux^{#1}_{#2}}
\newcommand\ljob[2]{\mathrm{#2}^{#1}_{}}
\newcommand\wjob{\widetilde{\mathrm{job}}}

\newcommand{\vin}[1][v]{#1^{\mathrm{in}}}
\newcommand{\vl}[1][v]{#1^{\mathrm{l}}}
\newcommand{\vr}[1][v]{#1^{\mathrm{r}}}
\newcommand{\vout}[1][v]{#1^{\mathrm{out}}}

\newcommand\fin{\mathrm{end}}
\newcommand\start{\mathrm{start}}
\newcommand\aux{\mathrm{aux}}

\usepackage[noend]{algpseudocode}
\algblock{Input}{EndInput}
\algnotext{EndInput}
\algblock{Output}{EndOutput}
\algnotext{EndOutput}

\newcommand{\bin}[2][2]{\binom{#2}{#1}}

\newcommand{\W}[1]{\ensuremath{\mathrm{W}[#1]}\xspace}
\newcommand\NP{\ensuremath{\mathrm{NP}}\xspace}
\newcommand\Ptime{\ensuremath{\mathrm{P}}\xspace}

\newcommand\FPT{\ensuremath{\mathrm{FPT}}\xspace}

\newcolumntype{R}[1]{>{\raggedleft\let\newline\\\arraybackslash\hspace{0pt}}m{#1}}

\newcommand{\taskprob}[3]{\prob[Task]{#1}{#2}{#3}}

\newcommand{\prob}[4][Question]{\begin{quote}  \textsc{#2}\\  \textbf{Input:} #3\\  \textbf{#1:} #4\end{quote}}

\newcommand{\mcc}{\textsc{Multicolored Clique}\xspace}
\newcommand{\SWAP}{\textsc{Swap LS MM}\xspace}
\newcommand{\INSERT}{\textsc{Insert LS MM}\xspace}
\newcommand{\TTMM}{\textsc{MM}\xspace}

\newcommand{\COMP}{CUNA\xspace}

\newcommand{\ND}{\mathrm{ND}}

\pgfdeclarelayer{background}
\pgfdeclarelayer{foreground}
\pgfsetlayers{background,main,foreground}

\newcommand{\Oh}{\mathcal{O}}

\newtheorem{myclaim}{Claim}{\bfseries}{\itshape}

\newcommand{\tableeqFive}{
\begin{tabular}{m{1.2cm}m{.6cm}|rrrrrr}
\hline
  &   & 50 & 100 & 150 & 200 & 250 & 300\\\hline\hline
Win & DD  & 735 & 1115 &  \cellcolor{black!10}1925 &  \cellcolor{black!10}4245 & 3835 & 5040\\
 & SM  & 715 & 1155 & 2965 & \dag &  \cellcolor{black!10}1450 &  \cellcolor{black!10}750\\
 & TM  & 725 & 1410 & 2260 & \dag & 1705 & 2630\\\hline
Win{\tiny +}Sw & DD  & 755 &  \cellcolor{black!10}1075 & 3195 & 4370 & 3365 & 4055\\
 & SM  & 715 & 1320 & 3010 & \dag & 2080 & 2015\\
 & TM  & 710 & 1335 & 2660 & \dag & 2125 & 1775\\\hline
MW & DD  & 1040 & 2605 & 4660 & 6430 & 8160 & 9955\\
 & SM  & 795 & 1645 & \dag & \dag & 2285 & \dag\\
 & TM  & 915 & 2060 & \dag & \dag & 3435 & 4325\\\hline
MW{\tiny +}Sw & DD  & 1000 & 1675 & 3955 & 5480 & 4625 & 7265\\
 & SM  & 970 & 1635 & 3140 & \dag & 2905 & 2075\\
 & TM  & 710 & 1840 & 3440 & \dag & 2270 & 2145\\\hline
PILS1 &  &  \cellcolor{black!10}705 & 1210 & 7455 & 19625 & 10175 & 9630\\
GAD &  & 3695 & 12970 & 18810 & \dag & 32850 & 46130\\
MGA &  & 1125 & 2810 & 5870 & 9540 & 10740 & 15250
\end{tabular} }

\newcommand{\tableeqSix}{
\begin{tabular}{m{1.2cm}m{.6cm}|rrrrrr}
\hline
  &   & 50 & 100 & 150 & 200 & 250 & 300\\\hline\hline
Win & DD  & 735 & 1115 &  \cellcolor{black!10}1925 &  \cellcolor{black!10}4245 & 3835 & 5040\\
 & SM  & 715 & 1155 & 2965 & \dag &  \cellcolor{black!10}1450 &  \cellcolor{black!10}750\\
 & TM  & 725 & 1410 & 2260 & \dag & 1705 & 2630\\\hline
Win{\tiny +}Sw & DD  & 755 &  \cellcolor{black!10}1075 & 3195 & 4370 & 3365 & 4055\\
 & SM  & 715 & 1320 & 3010 & \dag & 2080 & 2015\\
 & TM  & 710 & 1335 & 2660 & \dag & 2125 & 1775\\\hline
MW & DD  & 1040 & 2605 & 4660 & 6430 & 8160 & 9955\\
 & SM  & 795 & 1645 & \dag & \dag & 2285 & \dag\\
 & TM  & 915 & 2060 & \dag & \dag & 3435 & 4325\\\hline
MW{\tiny +}Sw & DD  & 1000 & 1675 & 3955 & 5480 & 4625 & 7265\\
 & SM  & 970 & 1635 & 3140 & \dag & 2905 & 2075\\
 & TM  & 710 & 1840 & 3440 & \dag & 2270 & 2145\\\hline
PILS1 &  &  \cellcolor{black!10}705 & 1210 & 7455 & 19625 & 10175 & 9630\\
GAD &  & 3695 & 12970 & 18810 & \dag & 32850 & 46130\\
MGA &  & 1125 & 2810 & 5870 & 9540 & 10740 & 15250\\\hline
\end{tabular} }

\newcommand{\tableeqLines}{
\begin{tabular}{m{1.2cm}m{.6cm}|rrrrrr}
\hline
  &   & 50 & 100 & 150 & 200 & 250 & 300\\\hline\hline
Win & DD  & 735 & 1115 &  \cellcolor{black!10}1925 &  \cellcolor{black!10}4245 & 3835 & 5040\\
 & SM  & 715 & 1155 & 2965 & \dag &  \cellcolor{black!10}1450 &  \cellcolor{black!10}750\\
 & TM  & 725 & 1410 & 2260 & \dag & 1705 & 2630\\\hline
Win{\tiny +}Sw & DD  & 755 &  \cellcolor{black!10}1075 & 3195 & 4370 & 3365 & 4055\\
 & SM  & 715 & 1320 & 3010 & \dag & 2080 & 2015\\
 & TM  & 710 & 1335 & 2660 & \dag & 2125 & 1775\\\hline
MW & DD  & 1040 & 2605 & 4660 & 6430 & 8160 & 9955\\
 & SM  & 795 & 1645 & \dag & \dag & 2285 & \dag\\
 & TM  & 915 & 2060 & \dag & \dag & 3435 & 4325\\\hline
MW{\tiny +}Sw & DD  & 1000 & 1675 & 3955 & 5480 & 4625 & 7265\\
 & SM  & 970 & 1635 & 3140 & \dag & 2905 & 2075\\
 & TM  & 710 & 1840 & 3440 & \dag & 2270 & 2145\\\hline
PILS1 &  &  \cellcolor{black!10}705 & 1210 & 7455 & 19625 & 10175 & 9630\\\hline
GAD &  & 3695 & 12970 & 18810 & \dag & 32850 & 46130\\\hline
MGA &  & 1125 & 2810 & 5870 & 9540 & 10740 & 15250
\end{tabular} }

\newcommand{\tableFirstZeroFive}{
\begin{tabular}{m{1.2cm}m{.6cm}|rrrrrr}
\hline
  &   & 50 & 100 & 150 & 200 & 250 & 300\\\hline\hline
Win & DD  &  \cellcolor{black!10}0 &  \cellcolor{black!10}0 &  \cellcolor{black!10}0 &  \cellcolor{black!10}0.188 &  \cellcolor{black!10}0 &  \cellcolor{black!10}0\\
 & SM  & 0.097 & 1.007 & 53.374 & \dag & 45.238 & 23.593\\
 & TM  & 0.054 & 0.794 & 28.907 & \dag & 23.329 & 67.621\\\hline
Win{\tiny +}Sw & DD  &  \cellcolor{black!10}0 &  \cellcolor{black!10}0 &  \cellcolor{black!10}0 & 1.317 &  \cellcolor{black!10}0 &  \cellcolor{black!10}0\\
 & SM  & 0.288 & 2.761 & 191.071 & \dag & 179.928 & 396.085\\
 & TM  & 0.015 & 2.023 & 99.118 & \dag & 27.204 & 179.501\\\hline
MW & DD  &  \cellcolor{black!10}0 &  \cellcolor{black!10}0 &  \cellcolor{black!10}0 & 6.308 &  \cellcolor{black!10}0 &  \cellcolor{black!10}0\\
 & SM  & 12.221 & 31.748 & \dag & \dag & 580.996 & \dag\\
 & TM  & 2.017 & 20.733 & \dag & \dag & 224.565 & 597.310\\\hline
MW{\tiny +}Sw & DD  &  \cellcolor{black!10}0 &  \cellcolor{black!10}0 &  \cellcolor{black!10}0 & 1.313 &  \cellcolor{black!10}0 &  \cellcolor{black!10}0\\
 & SM  & 0.148 & 2.741 & 184.951 & \dag & 174.482 & 396.164\\
 & TM  & 0.015 & 2.062 & 95.925 & \dag & 27.329 & 177.461\\\hline
PILS1 &  &  \cellcolor{black!10}0 &  \cellcolor{black!10}0 &  \cellcolor{black!10}0 & 502.505 &  \cellcolor{black!10}0 &  \cellcolor{black!10}0\\
GAD &  &  \cellcolor{black!10}0 &  \cellcolor{black!10}0 &  \cellcolor{black!10}0 & \dag &  \cellcolor{black!10}0 &  \cellcolor{black!10}0\\
MGA &  &  \cellcolor{black!10}0 &  \cellcolor{black!10}0 &  \cellcolor{black!10}0 & 48.115 &  \cellcolor{black!10}0 &  \cellcolor{black!10}0
\end{tabular} }

\newcommand{\tableFirstZeroSix}{
\begin{tabular}{m{1.2cm}m{.6cm}|rrrrrr}
\hline
  &   & 50 & 100 & 150 & 200 & 250 & 300\\\hline\hline
Win & DD  &  \cellcolor{black!10}0 &  \cellcolor{black!10}0 &  \cellcolor{black!10}0 &  \cellcolor{black!10}0.188 &  \cellcolor{black!10}0 &  \cellcolor{black!10}0\\
 & SM  & 0.097 & 1.007 & 53.374 & \dag & 45.238 & 23.593\\
 & TM  & 0.054 & 0.794 & 28.907 & \dag & 23.329 & 67.621\\\hline
Win{\tiny +}Sw & DD  &  \cellcolor{black!10}0 &  \cellcolor{black!10}0 &  \cellcolor{black!10}0 & 1.317 &  \cellcolor{black!10}0 &  \cellcolor{black!10}0\\
 & SM  & 0.288 & 2.761 & 191.071 & \dag & 179.928 & 396.085\\
 & TM  & 0.015 & 2.023 & 99.118 & \dag & 27.204 & 179.501\\\hline
MW & DD  &  \cellcolor{black!10}0 &  \cellcolor{black!10}0 &  \cellcolor{black!10}0 & 6.308 &  \cellcolor{black!10}0 &  \cellcolor{black!10}0\\
 & SM  & 12.221 & 31.748 & \dag & \dag & 580.996 & \dag\\
 & TM  & 2.017 & 20.733 & \dag & \dag & 224.565 & 597.310\\\hline
MW{\tiny +}Sw & DD  &  \cellcolor{black!10}0 &  \cellcolor{black!10}0 &  \cellcolor{black!10}0 & 1.313 &  \cellcolor{black!10}0 &  \cellcolor{black!10}0\\
 & SM  & 0.148 & 2.741 & 184.951 & \dag & 174.482 & 396.164\\
 & TM  & 0.015 & 2.062 & 95.925 & \dag & 27.329 & 177.461\\\hline
PILS1 &  &  \cellcolor{black!10}0 &  \cellcolor{black!10}0 &  \cellcolor{black!10}0 & 502.505 &  \cellcolor{black!10}0 &  \cellcolor{black!10}0\\
GAD &  &  \cellcolor{black!10}0 &  \cellcolor{black!10}0 &  \cellcolor{black!10}0 & \dag &  \cellcolor{black!10}0 &  \cellcolor{black!10}0\\
MGA &  &  \cellcolor{black!10}0 &  \cellcolor{black!10}0 &  \cellcolor{black!10}0 & 48.115 &  \cellcolor{black!10}0 &  \cellcolor{black!10}0\\\hline
\end{tabular} }

\newcommand{\tableFirstZeroLines}{
\begin{tabular}{m{1.2cm}m{.6cm}|rrrrrr}
\hline
  &   & 50 & 100 & 150 & 200 & 250 & 300\\\hline\hline
Win & DD  &  \cellcolor{black!10}0 &  \cellcolor{black!10}0 &  \cellcolor{black!10}0 &  \cellcolor{black!10}0.188 &  \cellcolor{black!10}0 &  \cellcolor{black!10}0\\
 & SM  & 0.097 & 1.007 & 53.374 & \dag & 45.238 & 23.593\\
 & TM  & 0.054 & 0.794 & 28.907 & \dag & 23.329 & 67.621\\\hline
Win{\tiny +}Sw & DD  &  \cellcolor{black!10}0 &  \cellcolor{black!10}0 &  \cellcolor{black!10}0 & 1.317 &  \cellcolor{black!10}0 &  \cellcolor{black!10}0\\
 & SM  & 0.288 & 2.761 & 191.071 & \dag & 179.928 & 396.085\\
 & TM  & 0.015 & 2.023 & 99.118 & \dag & 27.204 & 179.501\\\hline
MW & DD  &  \cellcolor{black!10}0 &  \cellcolor{black!10}0 &  \cellcolor{black!10}0 & 6.308 &  \cellcolor{black!10}0 &  \cellcolor{black!10}0\\
 & SM  & 12.221 & 31.748 & \dag & \dag & 580.996 & \dag\\
 & TM  & 2.017 & 20.733 & \dag & \dag & 224.565 & 597.310\\\hline
MW{\tiny +}Sw & DD  &  \cellcolor{black!10}0 &  \cellcolor{black!10}0 &  \cellcolor{black!10}0 & 1.313 &  \cellcolor{black!10}0 &  \cellcolor{black!10}0\\
 & SM  & 0.148 & 2.741 & 184.951 & \dag & 174.482 & 396.164\\
 & TM  & 0.015 & 2.062 & 95.925 & \dag & 27.329 & 177.461\\\hline
PILS1 &  &  \cellcolor{black!10}0 &  \cellcolor{black!10}0 &  \cellcolor{black!10}0 & 502.505 &  \cellcolor{black!10}0 &  \cellcolor{black!10}0\\\hline
GAD &  &  \cellcolor{black!10}0 &  \cellcolor{black!10}0 &  \cellcolor{black!10}0 & \dag &  \cellcolor{black!10}0 &  \cellcolor{black!10}0\\\hline
MGA &  &  \cellcolor{black!10}0 &  \cellcolor{black!10}0 &  \cellcolor{black!10}0 & 48.115 &  \cellcolor{black!10}0 &  \cellcolor{black!10}0
\end{tabular} }

\newcommand{\tablegeFive}{
\begin{tabular}{m{1.2cm}m{.6cm}|rrrrrr}
\hline
  &   & 50 & 100 & 150 & 200 & 250 & 300\\\hline\hline
Win & DD  & 889 & 8434 & 15213 & 586 & 15816 & 54422\\
 & SM  & 889 & 8420 & 15357 & 913 & 16781 & 54634\\
 & TM  & 890 & 8429 & 15178 & 773 & 16376 & 55052\\\hline
Win{\tiny +}Sw & DD  & 730 & 7296 &  \cellcolor{black!10}10774 & 505 &  \cellcolor{black!10}\emph{13772} &  \cellcolor{black!10}\emph{44903}\\
 & SM  & 731 &  \cellcolor{black!10}7251 & 10845 &  \cellcolor{black!10}475 & \emph{18784} & \emph{49399}\\
 & TM  & 735 & 7276 & 10806 & 491 & \emph{17365} & \emph{47011}\\\hline
MW & DD  & 953 & 8572 & 15682 & 649 & 16610 & 56935\\
 & SM  & 1000 & 8798 & 16315 & 1162 & 22320 & 59634\\
 & TM  & 910 & 8651 & 15733 & 897 & 19063 & 59019\\\hline
MW{\tiny +}Sw & DD  & 736 & 7323 & 10869 & 527 &  \cellcolor{black!10}\emph{13772} &  \cellcolor{black!10}\emph{44903}\\
 & SM  & 732 & 7308 & 10919 & 478 & \emph{18784} & \emph{49399}\\
 & TM  & 732 & 7316 & 10865 & 497 & \emph{17365} & \emph{47011}\\\hline
PILS1 &  &  \cellcolor{black!10}727 & 7675 & 12332 & 631 & 16316 & 48770\\
GAD &  & 810 & 8039 & 12823 & 1247 & 18512 & 48864\\
MGA &  & 904 & 8745 & 15975 & 699 & 17185 & 58540
\end{tabular} }

\newcommand{\tablegeSix}{
\begin{tabular}{m{1.2cm}m{.6cm}|rrrrrr}
\hline
  &   & 50 & 100 & 150 & 200 & 250 & 300\\\hline\hline
Win & DD  & 889 & 8434 & 15213 & 586 & 15816 & 54422\\
 & SM  & 889 & 8420 & 15357 & 913 & 16781 & 54634\\
 & TM  & 890 & 8429 & 15178 & 773 & 16376 & 55052\\\hline
Win{\tiny +}Sw & DD  & 730 & 7296 &  \cellcolor{black!10}10774 & 505 &  \cellcolor{black!10}\emph{13772} &  \cellcolor{black!10}\emph{44903}\\
 & SM  & 731 &  \cellcolor{black!10}7251 & 10845 &  \cellcolor{black!10}475 & \emph{18784} & \emph{49399}\\
 & TM  & 735 & 7276 & 10806 & 491 & \emph{17365} & \emph{47011}\\\hline
MW & DD  & 953 & 8572 & 15682 & 649 & 16610 & 56935\\
 & SM  & 1000 & 8798 & 16315 & 1162 & 22320 & 59634\\
 & TM  & 910 & 8651 & 15733 & 897 & 19063 & 59019\\\hline
MW{\tiny +}Sw & DD  & 736 & 7323 & 10869 & 527 &  \cellcolor{black!10}\emph{13772} &  \cellcolor{black!10}\emph{44903}\\
 & SM  & 732 & 7308 & 10919 & 478 & \emph{18784} & \emph{49399}\\
 & TM  & 732 & 7316 & 10865 & 497 & \emph{17365} & \emph{47011}\\\hline
PILS1 &  &  \cellcolor{black!10}727 & 7675 & 12332 & 631 & 16316 & 48770\\
GAD &  & 810 & 8039 & 12823 & 1247 & 18512 & 48864\\
MGA &  & 904 & 8745 & 15975 & 699 & 17185 & 58540\\\hline
\end{tabular} }

\newcommand{\tablegeLines}{
\begin{tabular}{m{1.2cm}m{.6cm}|rrrrrr}
\hline
  &   & 50 & 100 & 150 & 200 & 250 & 300\\\hline\hline
Win & DD  & 889 & 8434 & 15213 & 586 & 15816 & 54422\\
 & SM  & 889 & 8420 & 15357 & 913 & 16781 & 54634\\
 & TM  & 890 & 8429 & 15178 & 773 & 16376 & 55052\\\hline
Win{\tiny +}Sw & DD  & 730 & 7296 &  \cellcolor{black!10}10774 & 505 &  \cellcolor{black!10}\emph{13772} &  \cellcolor{black!10}\emph{44903}\\
 & SM  & 731 &  \cellcolor{black!10}7251 & 10845 &  \cellcolor{black!10}475 & \emph{18784} & \emph{49399}\\
 & TM  & 735 & 7276 & 10806 & 491 & \emph{17365} & \emph{47011}\\\hline
MW & DD  & 953 & 8572 & 15682 & 649 & 16610 & 56935\\
 & SM  & 1000 & 8798 & 16315 & 1162 & 22320 & 59634\\
 & TM  & 910 & 8651 & 15733 & 897 & 19063 & 59019\\\hline
MW{\tiny +}Sw & DD  & 736 & 7323 & 10869 & 527 &  \cellcolor{black!10}\emph{13772} &  \cellcolor{black!10}\emph{44903}\\
 & SM  & 732 & 7308 & 10919 & 478 & \emph{18784} & \emph{49399}\\
 & TM  & 732 & 7316 & 10865 & 497 & \emph{17365} & \emph{47011}\\\hline
PILS1 &  &  \cellcolor{black!10}727 & 7675 & 12332 & 631 & 16316 & 48770\\\hline
GAD &  & 810 & 8039 & 12823 & 1247 & 18512 & 48864\\\hline
MGA &  & 904 & 8745 & 15975 & 699 & 17185 & 58540
\end{tabular} }

\begin{document}


\title{Scalable Neighborhood Local Search for Single-Machine~Scheduling with Family Setup~Times}



\author[1]{Kaja Balzereit\footnote{kaja.balzereit@hsbi.de -- Supported by the German Federal Ministry for Education and Research (BMBF), Grant No.  03FHP106. The authors remain responsible for the content of this publication.}}
\author[2]{Niels Grüttemeier\footnote{\{niels.gruettemeier, stefan.windmann\}@iosb-ina.fraunhofer.de -- Supported by the German Federal Ministry for the Environment, Nature Conservation, Nuclear Safety and Consumer Protection (BMUV) under the project \emph{Smart-E-Factory}.}}
\author[3]{Nils Morawietz\footnote{nils.morawietz@uni-jena.de}}
\author[4]{Dennis Reinhardt\footnote{Dennis.Reinhardt@sollich.com}}
\author[2]{Stefan Windmann$^\dagger$}
\author[5]{Petra Wolf\footnote{mail@wolfp.net -- Supported by the French ANR, project ANR-22-CE48-0001 (TEMPOGRAL).}}

\affil[1]{\small Hochschule Bielefeld -- University of Applied Sciences and Arts, Bielefeld, Germany}

\affil[2]{\small Fraunhofer IOSB-INA, Lemgo, Germany}

\affil[3]{\small Friedrich Schiller University Jena, Institute of Computer Science, Jena, Germany}

\affil[5]{\small Sollich KG, Bad Salzuflen, Germany}

\affil[5]{\small LaBRI, CNRS, Université de Bordeaux, Bordeaux INP, France}

\maketitle              
%


\begin{abstract}
In this work, we study the task of scheduling jobs on a single machine with sequence dependent family setup times under the goal of minimizing the makespan.
This notoriously NP-hard problem is highly relevant in real-world industrial productions and requires heuristics that provide good solutions quickly in order to deal with large instances.
In this paper, we present a heuristic based on the approach of parameterized local search. 
That is, we aim to replace a given solution by a better solution having distance at most~$k$ in a pre-defined distance measure. 
This is done multiple times in a hill-climbing manner, until a locally optimal solution is reached. 
We analyze the trade-off between the allowed distance~$k$ and the algorithm's running time for four natural distance measures. Examples of allowed operations for our considered distance measures are: swapping~$k$ pairs of jobs in the sequence, or rearranging~$k$ consecutive jobs.
For two distance measures, we show that finding an improvement for a given~$k$ can be done in~$f(k) \cdot n^{\Oh(1)}$~time, while such a running time for the other two distance measures is unlikely.
We provide a preliminary experimental evaluation of our local search approaches. 

\end{abstract}

\todomi{add fundings (?)}
\todomi{add emails}
\todomi{paragraphs might look weird...}



\section{Introduction}
Finding orderings in which products are manufactured on a machine is among the most important problems in combinatorial optimization and is
highly relevant in real-world industrial production. 
It has applications in many real-world scenarios appearing for instance in the solar cell industry~\cite{C13} or the cider industry~\cite{RNS018}. 
Since the mid-1960s, scheduling problems are extensively studied for a wide range of shop floor models, target functions, and setup information~\cite{A15}. 

In this work, we study the task of scheduling jobs on a single machine with sequence dependent family-setup times~\cite{GBSB23,TB12} under the goal of minimizing the makespan, that is, the completion time of the last job in the schedule. 
In this problem, one is given a set of jobs, each with an individual processing time, a deadline, and some identifier for a product type. 
Additionally, one is given a setup matrix specifying the additional time it takes to change the machine setup from one product type to another. 
The goal is to find a sequence of all jobs such that every job meets its deadline and the completion time of the last job is minimal.  
Our work is motivated by the real-world production scheduling of the company~\COMP. 
\COMP manufactures products that come in eight different material/color combinations on a single injection molding machine and changing the setup from one material/color combination to another takes a combination specific amount of time. 

The considered scheduling problem is \NP-hard as it obviously generalizes the famous \textsc{Traveling Salesperson Problem (TSP)}. 
Due to this intractability, instances with a large number of jobs are usually solved using heuristics. 
A common approach to tackle scheduling problems are genetic algorithms~\cite{OX14,LJ07}.
Since the applications of genetic algorithms usually have a large number of parameters that play an important role in their performance~\cite{CD09}, it can be very difficult to reconstruct studied algorithms~\cite{RNS21} and make them produce equally good results for a new use case. 
Moreover, genetic algorithms usually do not provide a (local) optimality guarantee for the returned solution. 
Another important class of heuristics are local search algorithms, which usually have a simple design and provide a local optimality guarantee on the returned solution: In a hill climbing manner, one aims to improve a solution by performing \emph{one} small change on a current solution, until no further improvement can be found. 
Variable Neighborhood Search (VNS) strategies that aim to find an improvement by combining swapping two jobs and inserting a job at another position in the schedule~\cite{EJBR09,RSA12} are considered a state-of-the-art approach for single-machine scheduling
~\cite{A15}.

Our approach is based on another version of local search called \emph{parameterized local search}. Parameterized local search combines the paradigms of parameterized algorithmics~\cite{C+15} with local search heuristics. 
While in other approaches a solution is improved by \emph{one} operation~\cite{EJBR09,RSA12}, the user may set a search radius~$k$ and extend the search space for possible improvements with up to~$k$ operations. More precisely, one aims to find an improving solution within a radius of size~$k$ around the current solution for some distance function. This technique might prevent getting stuck in bad local optima. 
With this work, we aim to outline the trade-off between the running time of the heuristic and the size of the search radius~$k$. 
Parameterized local search lately received much attention in the algorithmics community both from a theoretical~\cite{BIJK19,DGKW14,GKO+12,GHNS13,KLMS23,Szei11} and practical~\cite{GGJ+19,GGKM23,GKM21,HN13,KK17} point of view. In case of sequencing problems, parameterized local search has been studied for TSP~\cite{M08} and for finding topological orderings of Bayesian networks~\cite{GKM21}. 

In the realm of parameterized local search, the choice of the distance function on the solution space and the choice of the search radius~$k$ in which we aim to find an improvement is crucial. We now discuss the four distances investigated in this work.
When considering the \emph{insert distance}, one aims to improve the solution by removing up to~$k$ jobs from the ordering and inserting them at new positions.
When considering the \emph{swap distance}, one aims to improve the solution by sequentially swapping up to~$k$ pairs of jobs in a given schedule. 
Note that parameterized local search for these two distances essentially lifts established local search strategies via single swaps and single inserts~\cite{EJBR09,RSA12} to the parameterized approach and is thus a natural next step in the design of hill-climbing strategies.
When considering the \emph{window distance}, one checks whether there exists a window consisting of~$k$ consecutive jobs in a given schedule, such that the solution can be improved by rearranging the jobs inside the window. 
When considering the \emph{multi-window distance}, one can rearrange the jobs in arbitrarily many disjoint windows, of up to~$k$ consecutive jobs each, simultaneously to obtain an improved solution. Figure~\ref{Figure: Example neighborhoods} shows one example of each studied distance. 
Note that the multi-window distance can be seen as an extension of the window distance in the following sense: The set of possible schedules with multi-window distance~$k$ is a superset of the schedules with window distance~$k$ from a given schedule. 
The multi-window distance is closely related to a multi-inversions distance used for finding Bayesian network structures via parameterized local search~\cite{GKM21}.
Note that for all these distances, the problem of searching for a better schedule with distances at most~$k$ becomes~\NP-hard, since for sufficiently large~$k$, one aims to find a globally optimal solution. 

\paragraph{Related Work.}Parameterized algorithmics for single machine scheduling has been studied extensively: The parameterized complexity of minimizing the weighted number of tardy jobs has been studied~\cite{HKPS21}, also in combination with the weighted completion time~\cite{HHS23}. Fellows and McCartin~\cite{FM03} studied single-machine scheduling with precedence constraints. For the scenario where jobs can be rejected with some rejection cost, de Weerdt et al.~\cite{WBH21} provided results on the parameterized algorithmics and studied approximation schemes. A relatively new setting is equitable single-machine scheduling with multiple agents guaranteeing each agent a minimal level of service~\cite{HHMMNS23}.
For a structured overview, especially on open challenges in parameterized algorithmics for scheduling problems, we refer to the survey by Mnich and van Bevern~\cite{MB18}.
Our single-machine scheduling problem with the \COMP application has previously been studied in a purely practical work~\cite{GBSB23}. A problem version with unit processing times has been studied by Cheng and Kovalyov~\cite{CK01}. Rego et al.~\cite{RSA12} proposed an efficient hill-climbing heuristic for the problem.

Parameterized local search for sequencing problems has been studied for TSP~\cite{M08} and for finding topological orderings of Bayesian networks~\cite{GKM21}. Parameterized local search for industrial production planning has previously been studied for \textsc{Vector Bin Packing} and other partitioning problems~\cite{GMS25}.
 This work is---to the best of our knowledge---the first work studying parameterized local search for a scheduling~problem.
 

\def\myOffset{.8}
\begin{figure}[t]
\begin{center}
\scalebox{01}{
\begin{tikzpicture}[scale=0.85,yscale=0.6,xscale=.9]
\tikzstyle{knoten}=[rectangle,fill=white,draw=black,minimum height=6pt,minimum width=10pt,inner sep=0pt]
\tikzstyle{bez}=[inner sep=0pt]

\begin{scope}[yshift=.5cm]
\node[knoten,label=below:$1$] (v1) at (-10,10) {};
\node[knoten,label=below:$2$] (v2) at ($(v1) + (\myOffset,0)$) {};
\node[knoten,label=below:$3$] (v3) at ($(v2) + (\myOffset,0)$) {};
\node[knoten,label=below:$4$] (v4) at ($(v3) + (\myOffset,0)$) {};
\node[knoten,label=below:$5$] (v5) at ($(v4) + (\myOffset,0)$) {};
\node[knoten,label=below:$6$] (v6) at ($(v5) + (\myOffset,0)$) {};
\node[knoten,label=below:$7$] (v7) at ($(v6) + (\myOffset,0)$) {};
\node[knoten,label=below:$8$] (v8) at ($(v7) + (\myOffset,0)$) {};

\node (label) at ($(v1) - (\myOffset,0)$) {$\pi$};

\end{scope}

\begin{scope}[yshift=-2.5cm,xshift=4cm]
\node[knoten,label=below:$1$] (v1) at (-10,10) {};
\node[knoten,label=below:$2$] (v2) at ($(v1) + (\myOffset,0)$) {};
\node[knoten,label=below:$\color{red}4$] (v3) at ($(v2) + (\myOffset,0)$) {};
\node[knoten,label=below:$\color{red}6$] (v4) at ($(v3) + (\myOffset,0)$) {};
\node[knoten,label=below:$5$] (v5) at ($(v4) + (\myOffset,0)$) {};
\node[knoten,label=below:$\color{red}3$] (v6) at ($(v5) + (\myOffset,0)$) {};
\node[knoten,label=below:$7$] (v7) at ($(v6) + (\myOffset,0)$) {};
\node[knoten,label=below:$8$] (v8) at ($(v7) + (\myOffset,0)$) {};
\node (label) at ($(v1) - (\myOffset,0)$) {$\pi_3$};

\begin{pgfonlayer}{background}
\draw[rounded corners, fill=black!10, draw=black!10] ($(v3) + (-.4,-.4)$) rectangle ($(v6) + (.4,.4)$) {};
\end{pgfonlayer}
\end{scope}

\begin{scope}[yshift=-4.5cm,xshift=4cm]
\node[knoten,label=below:$1$] (v1) at (-10,10) {};
\node[knoten,label=below:$\color{red}4$] (v2) at ($(v1) + (\myOffset,0)$) {};
\node[knoten,label=below:$\color{red}2$] (v3) at ($(v2) + (\myOffset,0)$) {};
\node[knoten,label=below:$\color{red}3$] (v4) at ($(v3) + (\myOffset,0)$) {};
\node[knoten,label=below:$\color{red}7$] (v5) at ($(v4) + (\myOffset,0)$) {};
\node[knoten,label=below:$6$] (v6) at ($(v5) + (\myOffset,0)$) {};
\node[knoten,label=below:$\color{red}5$] (v7) at ($(v6) + (\myOffset,0)$) {};
\node[knoten,label=below:$8$] (v8) at ($(v7) + (\myOffset,0)$) {};

\node (label) at ($(v1) - (\myOffset,0)$) {$\pi_4$};

\begin{pgfonlayer}{background}
\draw[rounded corners, fill=black!10, draw=black!10] ($(v2) + (-.4,-.4)$) rectangle ($(v4) + (.4,.4)$) {};
\draw[rounded corners, fill=black!10, draw=black!10] ($(v5) + (-.4,-.4)$) rectangle ($(v7) + (.4,.4)$) {};
\end{pgfonlayer}
\end{scope}

\begin{scope}[yshift=-2.5cm,xshift=-4cm]
\node[knoten,label=below:$1$] (v1) at (-10,10) {};
\node[knoten,label=below:$2$] (v2) at ($(v1) + (\myOffset,0)$) {};
\node[knoten,label=below:$\color{red}6$] (v3) at ($(v2) + (\myOffset,0)$) {};
\node[knoten,label=below:$\color{red}3$] (v4) at ($(v3) + (\myOffset,0)$) {};
\node[knoten,label=below:$\color{red}4$] (v5) at ($(v4) + (\myOffset,0)$) {};
\node[knoten,label=below:$\color{red}5$] (v6) at ($(v5) + (\myOffset,0)$) {};
\node[knoten,label=below:$7$] (v7) at ($(v6) + (\myOffset,0)$) {};
\node[knoten,label=below:$8$] (v8) at ($(v7) + (\myOffset,0)$) {};
\node (label) at ($(v1) - (\myOffset,0)$) {$\pi_1$};

\draw[<-, dashed, line width=1pt, bend left=40]  (v3) to ($.5*(v6) + .5*(v7) + (0,.2)$);
\end{scope}

\begin{scope}[yshift=-4.5cm,xshift=-4cm]
\node[knoten,label=below:$1$] (v1) at (-10,10) {};
\node[knoten,label=below:$\color{red}7$] (v2) at ($(v1) + (\myOffset,0)$) {};
\node[knoten,label=below:$3$] (v3) at ($(v2) + (\myOffset,0)$) {};
\node[knoten,label=below:$\color{red}8$] (v4) at ($(v3) + (\myOffset,0)$) {};
\node[knoten,label=below:$\color{red}6$] (v5) at ($(v4) + (\myOffset,0)$) {};
\node[knoten,label=below:$\color{red}5$] (v6) at ($(v5) + (\myOffset,0)$) {};
\node[knoten,label=below:$\color{red}2$] (v7) at ($(v6) + (\myOffset,0)$) {};
\node[knoten,label=below:$\color{red}4$] (v8) at ($(v7) + (\myOffset,0)$) {};

\draw[<->, dashed, line width=1pt, bend left=40]  ($(v5) - .05*(\myOffset,-4)$) to ($(v6) + .05*(\myOffset,4)$);
\draw[<->, dashed, line width=1pt, bend left=40]  (v4) to (v8);
\draw[<->, dashed, line width=1pt, bend left=40]  (v2) to (v7);
\node (label) at ($(v1) - (\myOffset,0)$) {$\pi_2$};

\end{scope}

\end{tikzpicture}
}
\end{center}
\caption{Examples of the considered distance measures:
the insert distance of~$\pi$ and~$\pi_1$ is~1,
the swap distance of~$\pi$ and~$\pi_2$ is~3, 
the window distance of~$\pi$ and~$\pi_3$ is~4, and
the multi-window distance of~$\pi$ and~$\pi_4$ is~3.}\label{Figure: Example neighborhoods}
\end{figure}
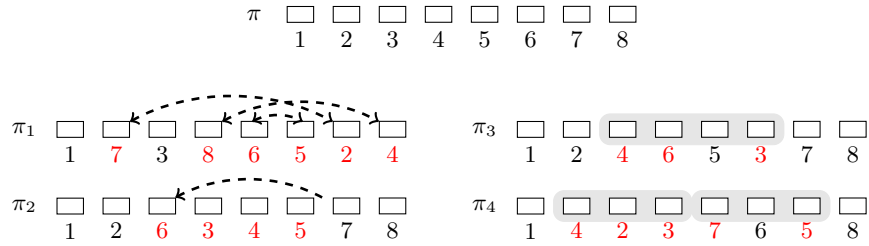

\paragraph{Our Contribution.} 
We initiate the study of parameterized local search in the context of single machine scheduling.
We outline to which extent this approach yields promising results for the study of makespan minimization.
We study four natural local search distances introduced above, analyze the parameterized complexity of the corresponding local search problems when parameterized by the search radius $k$, and evaluate our findings experimentally for the real-world setup matrix from company~\COMP{}'s injection molding machine.

We show
for the window distance and the multi-window distance that one can find in time~$k^{\min{(k,t)}} \cdot |I|^{\Oh(1)}$ an improving schedule if one exists. 
Herein,~$|I|$ denotes the total input size and~$t$ denotes the number of distinct product types in the instance. 
Note that in this running time, there is no exponential dependence on the total input size.
Since the number of different product types manufactured on one machine is usually bounded by a small constant, the factor~$k^{\min{(k,t)}}$ provides a good trade-off between the running time and the size of the search radius~$k$. 
We complement this result by showing that for the insert and the swap distance, there is no algorithm with running time~$f(k) \cdot |I|^{\Oh(1)}$ for any computable function~$f$, unless~$\W{1}=\FPT$. 
%
While the contributions of this paper are mainly theoretical results for the parameterized local search problems, we also provide preliminary experiments to asses whether parameterized local search is in principle a viable approach in practice. More precisely, we evaluate the solution quality of our hill-climbing approaches on 12 test instances. Motivated by the practical relevance of genetic algorithms and by the good results of VNS strategies, we consider two genetic algorithms and a state-of-the-art hill-climbing algorithm PILS1~\cite{RSA12} as a base line for the comparison of our preliminary experiments.
%


\section{Preliminaries}
Given some integers~$a \in \mathds{N}_0$ and~$b \in \mathds{N}_0$, we let~$[a,b]:=\{n \in \mathds{N}_0 \mid a \leq n \leq b\}$. We study a problem of sequencing jobs on a single machine with sequence-dependent family setup times. Let~$\mathcal{T}$ be an arbitrary finite set of \emph{product types}.
\begin{definition}
A \emph{job} is a triple~$j:=(t_j,d_j,\tau_j)$. Herein, the value~$t_j \in \mathds{N}_0$ denotes the processing time of~$j$, the value~$d_j \in \mathds{N}_0$ denotes the deadline of~$j$, and~$\tau_j \in \mathcal{T}$ is the type of~$j$.
\end{definition}

\begin{definition}
A \emph{setup mapping} is a mapping~$\ell: \mathcal{T} \times \mathcal{T} \rightarrow \mathds{N}_0$.
\end{definition}

In context of our application, given two types~$\tau$ and~$\tau'$, the value~$\ell(\tau,\tau')$ is the time needed to set up the machine if a job of type~$\tau'$ is scheduled right after a job of type~$\tau$. We assume that~$\ell(\tau, \tau) = 0$ for every~$\tau \in \mathcal{T}$ and that~$\ell$ satisfies the triangle inequality. That is,~$\ell(\tau_1,\tau_3) \leq \ell(\tau_1,\tau_2) + \ell(\tau_2,\tau_3)$ for every triple~$(\tau_1, \tau_2, \tau_3)$ of types.

\begin{definition}
Let~$J:=\{j_1, \dots, j_n\}$. A \emph{schedule} is a permutation~$\pi$ of~$J$. Given a schedule~$\pi$ of~$J$, we define the \emph{completion time of the~$i$th job on~$\pi$} as
$$
C^J_\pi(i) :=
\begin{cases}
t_{\pi(1)} & \text{if }i=1,\\
C^J_\pi(i-1) + \ell(\tau_{\pi(i-1)},\tau_{\pi(i)})+t_{\pi(i)} & \text{if }i>1.
\end{cases}
$$
We may omit the superscript~$J$ if the job set is clear from the context. The \emph{makespan}~$C^\pi_{\max}$ is then defined as the completion time of the last job on~$\pi$. We may write~$C_{\max}$ for the makespan if~$\pi$ is clear from the context.
\end{definition}

Furthermore, we define the \emph{total tardiness of~$\pi$} as
\iflong
$$
\sum_{i=1}^n \max(C_\pi (i) - d_{\pi(i)}, 0).
$$\else
$\sum_{i=1}^n \max(C_\pi (i) - d_{\pi(i)}, 0).$
\fi
Throughout this work, we call a schedule~$\pi$ \emph{feasible} if the total tardiness of~$\pi$ is~$0$. Intuitively, a schedule is feasible if every job meets its deadline. Motivated by our practical use case, we aim to find feasible schedules with minimum makespan. 

\taskprob{Makespan Minimization (MM)}
{A set~$J$ of jobs and a setup mapping~$\ell$.}
{Find a feasible schedule~$\pi$ of~$J$ that minimizes~$C_{\max}$.}

In the systematic \emph{Graham notation}~\cite{G79}, MM is the problem~$1 \mid \text{ST}_\text{sd,f}, \overline{d}_j \mid C_{\max}$\iflong{}: Scheduling on one machine, when we have  sequence-dependent family setup times, strict deadlines, and aiming to minimize the makespan~$C_{\max}$\fi{}.
\iflong

\fi
Throughout this work, we let~$n$ denote the number of jobs in an instance~$(J,\ell)$ of MM. Given two indices~$a$, and~$b$, we write~$a <_\pi b$ if~$a$ precedes~$b$ in a sequence~$\pi$. We let~$\pi[i,j]$ denote the subsequence~$(\pi(i), \pi(i+1), \dots,\pi(j))$. Furthermore, we let~$\pi \circ \sigma$ denote the concatenation of the sequences~$\pi$ and~$\sigma$.

\paragraph{Parameterized Local Search.} A \emph{distance measure}~$\delta: \pi \times \pi' \mapsto x \in \mathds{N}_0$ maps a pair of schedules to a non-negative integer. The distance measures considered in this work satisfy~$\delta(\pi,\pi)=0$ for every~$\pi$ and are symmetric, that is,~$\delta(\pi,\pi')=\delta(\pi',\pi)$ for every~$\pi$ and~$\pi'$. 

In our heuristics for MM, we are given a schedule~$\pi$ and some integer~$k$, and we aim to compute a better schedule that has distance at most~$k$ with~$\pi$ for some distance measure~$\delta$. 
Formally, we solve the following computational problem.

\taskprob{$\delta$ Local Search Makespan Minimization ($\delta$ LS MM)}
{A set~$J$ of jobs, a setup mapping~$\ell$, a feasible schedule~$\pi$ of~$J$, and an integer~$k$.}{Find a schedule~$\pi'$ \iflong{}of~$J$ \fi{}with~$\delta(\pi,\pi')\leq k$ such that~$\pi'$ is \emph{better} than~$\pi$, that is, $\pi'$ is feasible and $C_{\max}^{\pi'} < C_{\max}^{\pi}$. Or report that no better schedule exists.}

Recall that Figure~\ref{Figure: Example neighborhoods} provides an intuitive illustration for the four distance measures that we study in this work.

Let~$\delta$ be a distance measure. 
We say that~\textsc{$\delta$ LS MM} is \emph{fixed-parameter tractable (\FPT) for~$k$} if it can be solved in~$f(k) \cdot |I|^{\Oh(1)}$~time for some function~$f$. 
That is, the whole size~$|I|$ of the instance only contributes as a polynomial factor to the running time, while the exponential factor in the running time only depends on~$k$. 
Thus, the running time of~$f(k) \cdot |I|^{\Oh(1)}$ nicely outlines the trade-off between solution quality (radius size) and running time of the heuristic. 
If a problem is~\W1-hard when parameterized by~$k$ it is assumed that it is not fixed-parameter tractable for~$k$. 
For a detailed introduction into parameterized complexity, we refer to the standard monograph~\cite{C+15}.

\section{Local Search in Window Neighborhoods}\label{sec:3}
In this section, we show that the local search problems~\textsc{Window LS MM} and~\textsc{Multi-Window LS MM}, corresponding to the problem $\delta$ LS MM when the window neighborhood and multi-window neighborhood is concerned, are both FPT when parameterized by~$k$. The section is structured as follows. First, we formally introduce the distance measures leading to the local search problems~\textsc{Window LS MM} and~\textsc{Multi-Window LS MM} and we prove an observation on the solution structure of these problems, that we will later exploit in our algorithms. Second, we provide an efficient subroutine for improving windows of a schedule. Third, we describe how this subroutine can be used to obtain FPT algorithms for both local search problems.


\paragraph{Problem Definitions.} We first consider the window distance. The intuition behind the window distance is that one permutation results from another by changing the ordering of jobs inside one window (see~$\pi_3$ in Figure~\ref{Figure: Example neighborhoods}). Formally, let~$\pi$ and~$\pi'$ be permutations of~$n$ jobs. If~$\pi=\pi'$, the schedules have window distance~$0$. Otherwise, the \emph{window distance} is~$b-a+1$, where~$a$ ($b$) is the smallest (largest) index~$i$ of~$[1,n]$ with~$\pi(i) \neq \pi'(i)$.
  In the remainder of this work, we let~\textsc{Window LS MM} denote the local search problem~\textsc{$\delta$ LS MM}, where~$\delta$ is the window distance.

Given the schedules~$\pi$ and~$\pi'$ and the index set~$[a,b]$ from the previous definition, we call the subsequence~$\pi'[a,b]$ the \emph{rearranged window}.

We next define the multi-window distance. The intuition behind the multi-window distance is to split a permutation in arbitrary many intervals of bounded length and only rearrange the jobs inside the intervals~(see~$\pi_4$ in Figure~\ref{Figure: Example neighborhoods}). Formally, let~$\pi$ and~$\pi'$ be schedules for the job set~$J$. If~$\pi=\pi'$, the schedules have multi-window distance~$0$. Otherwise, we say that~$\pi$ and~$\pi'$ \emph{decompose into~$k$-intervals} for some~$k \in \mathds{N}$,~if
\begin{enumerate}
\item[$a)$] $\pi$ and~$\pi'$ have length at most~$k$, or
\item[$b)$] there is some index~$i \in [n-k,n-1]$ such that~$\{\pi(t) \mid t \in [i+1,n]\}= \{\pi'(t) \mid t \in [i+1,n]\}$, and the subsequences~$\pi[1,i]$ and~$\pi'[1,i]$ decompose into~$k$-intervals.
\end{enumerate}
Note that the decomposition into~$k$-intervals is well-defined as~$\{\pi(t) \mid t \in [i+1,n]\}= \{\pi'(t) \mid t \in [i+1,n]\}$ implies that the subsequences~$\pi[1,i]$ and~$\pi'[1,i]$ schedule the same jobs. The \emph{multi-window distance of~$\pi$ and~$\pi'$} is then defined as the minimal value~$k$ for which~$\pi$ and~$\pi'$ decompose into~$k$-intervals. In the remainder of this work, we let~\textsc{Multi-Window LS MM} denote the local search problem~\textsc{$\delta$ LS MM}, where~$\delta$ is the multi-window distance.

Given two schedules~$\pi$ and~$\pi'$ that have multi-window distance at most~$k$, the indices~$i$ from the recursive definition above introduce a (not necessary unique) partition of~$[1,n]$ into intervals~$\{I_1, \dots, I_m\}$. For one fixed partition, we call the subsequences~$\pi'[I_i]$ the \emph{rearranged windows}.

Observe that for some schedule~$\pi$ there may exist an improving schedule~$\pi'$ that has~multi-window distance~$k$ from~$\pi$, while there is no improvement within window distance at most~$k$.
\iflong{}Consider the following simple example: We have types~$\mathcal{T}=\{1,2,3\}$ and a setup mapping~$\ell$ with~$\ell (\tau_1,\tau_2)=0$ if~$\tau_1=\tau_2$, and~$\ell (\tau_1,\tau_2)=1$ otherwise. Moreover, we have four jobs~$j_1, j_2, j_3, j_4$, each with processing time one, deadline~$\infty$, and the types are~$1,2,3,1$, respectively. Given a schedule~$\pi=(j_1,j_2,j_3,j_4)$, the schedule~$\pi'=(j_2,j_1,j_4,j_3)$ has a strictly smaller makespan and multi window distance~$k=2$ from~$\pi$. In contrast, there is no improving schedule with window distance~$k=2$ from~$\pi$. \todog{Streichpotential?}\fi


\paragraph{An Observation on the Solution Structure.}
Recall that the goal of this section is to provide efficient FPT algorithms for~\textsc{Window LS MM} and for~\textsc{Multi-Window LS MM}. This is done by adapting a dynamic programming algorithm that exploits an observation on the solution structure: We may limit our search space to solutions, where the jobs of each type are sorted by their deadlines.

A similar algorithm has previously been used by Cheng and Kovalyov~\cite{CK01} to solve a special case of~MM where all jobs of each type have the same processing time. Grüttemeier et al.~\cite{GBSB23} also used this algorithmic idea to compute exact solutions for small instances of MM with an unbounded number of processing times per type. However, the proof of the observation was omitted in their work.
In this work, we provide the observation for the more general local search problems \textsc{Window LS MM} and for~\textsc{Multi-Window LS MM} with an arbitrary number of different processing times per type. 

%

To formally describe and prove the observation, we first introduce some terminology: 
\begin{definition}
Let~$(J,\ell)$ be an instance of MM with types in~$\mathcal{T}$. Furthermore, let~$\tau \in \mathcal{T}$, and let~$\pi$ be a schedule for~$(J,\ell)$. 
\begin{enumerate}
\item[•] The \emph{type-induced schedule} of~$\pi$ and~$\tau$ is the subsequence~$\pi[\tau]$ of~$\pi$ containing exactly the jobs of type~$\tau$.
\item[•] A schedule~$\pi$ is called \emph{earliest due date schedule (EDDS)} if for every~$\tau \in \mathcal{T}$, the type-induced schedule~$\pi[\tau]$ is a schedule of all jobs of type~$\tau$ in non-decreasing manner by their deadlines.
\item[•] A pair~$(a,b)$ of indices is called~\emph{type-inversion on~$\pi$}, if~$a <_\pi b$, the corresponding jobs~$j_a$ and~$j_b$ have the same type, and~$d_{j_a} > d_{j_b}$.
\end{enumerate}
\end{definition}
Note that~$\pi$ is an EDDS if and only if~$\pi$ has no type-inversions.

\begin{proposition} \label{Prop: EDDS}
Let~$\ell$ be a setup mapping such that~$\ell(\tau,\tau)=0$ for every type~$\tau$, and~$\ell$ satisfies the triangle inequality. 
\iflong
Moreover, let~$\delta$ be the distance measure for the window distance or the multi-window distance.

If~$(J,\ell,\pi,k)$ is an instance of \textsc{$\delta$ LS MM} such that there exists an improving feasible schedule with distance at most~$k$, then there is an improving feasible schedule~$\pi'$ with distance at most~$k$, such that
\begin{enumerate}
\item[$a)$] the rearranged window~$\pi'[a,b]$ is an EDDS in case of window distance, and
\item[$b)$] all rearranged windows~$\pi'[I_i]$ are EDDS in case of multi-window distance.
\end{enumerate} 
\else
If~$(J,\ell,\pi,k)$ is an instance of \textsc{Window LS MM} or \textsc{Multi-Window LS MM} such that there is a better schedule with distance at most~$k$, then there is a better schedule~$\pi'$ with distance at most~$k$, such that each rearranged window is an EDDS.
\fi
%
\end{proposition}

\newcommand{\proofOfObservation}{

\iflong
\begin{proof}
\else
\begin{proof}[Proof of~\Cref{Prop: EDDS}]
\fi

\iflong
We prove the proposition by first showing Statement~$b)$. Afterwards, we argue how the same arguments can be used to also show Statement~$a)$. 
\else
We prove the proposition by first considering \textsc{Multi-Window LS MM}. Afterwards, we argue how the same arguments can be used to also show Statement for \textsc{Window LS MM}. 
\fi
Let~$\pi'$ be a feasible schedule with multi-window distance at most~$k$ from~$\pi$ such that~$C^{\pi'}_{\max} < C^{\pi}_{\max}$. Furthermore, let~$\xi$ be the total number of type-inversions inside all rearranged windows~$\pi'[I_i]$. Without loss of generality, we assume that~$\pi'$ is chosen in a way that~$\xi$ is minimal. We prove the statement by showing~$\xi=0$.

Assume towards a contradiction that~$\xi>0$. Then, there is one rearranged window~$\pi'[I_i]$ containing a pair of jobs~$j_a <_{\pi'[I_i]} j_b$, such that~$j_a$ and~$j_b$ have the same type~$\tau \in \mathcal{T}$ and~$d_{j_a} > d_{j_b}$. We change the job schedule inside~$\pi'[I_i]$ in a way that the number of type-inversions on~$\pi'[I_i]$ becomes strictly smaller, while not violating any deadline  and not increasing~$C^{\pi'}_{\max}$. Note that this does also not increase the multi-window distance between~$\pi$ and~$\pi'$. Therefore, such transformation contradicts the minimality of~$\xi$.

Let~$\sigma$ be the subsequence of~$\pi'[I_i]$ that contains all jobs between~$j_a$ and~$j_b$ on~$\pi'$. We consider the case where~$\sigma$ is the empty sequence and the case where~$\sigma$ is non-empty separately.

\textbf{Case 1:} $\sigma$ is the empty sequence\textbf{.} That is, no job is scheduled between~$j_a$ and~$j_b$. We replace~$\pi'$ by~$\pi'_\text{new}$, which we define as
$\pi'_\text{new} := \pi'_\text{pre} \circ (j_b,j_a) \circ \pi'_{\text{suf}}.$
Herein,~$\pi'_\text{pre}$, $\pi'_{\text{suf}}$ denote the (possibly empty) prefix and suffix of~$\pi'$ containing the jobs scheduled before and after~$(j_a,j_b)$. Intuitively,~$\pi'_\text{new}$ results from~$\pi'$ by swapping~$j_a$ and~$j_b$. Since we only rearranged jobs inside~$\pi'[I_i]$, the resulting schedule~$\pi'_\text{new}$ has multi-window distance at most~$k$ from~$\pi$.

Note that~$(j_b,j_a)$ is not a type-inversion on~$\pi'_\text{new}$, while for every other pair~$(x,y) \neq (j_a,j_b)$ we have~$x <_{\pi'_\text{new}} y$ if and only if~$x <_{\pi'} y$. Consequently, the number of type-inversions on~$\pi'_\text{new}[I_i]$ is smaller than on~$\pi'[I_i]$. Moreover, since both jobs have the same type~$\tau$ and~$\ell(\tau,\tau)=0$, swapping the jobs does not increase the sum of setup times and all jobs on~$\pi'_\text{pre}$ and~$\pi'_{\text{suf}}$ have the same completion time on~$\pi'$ and~$\pi'_\text{new}$.

It remains to show that~$j_a$ and~$j_b$ meet their deadlines on~$\pi'_\text{new}$. Let~$c_b$ denote the completion time of~$j_b$ under~$\pi'$. By the fact that~$\pi'$ is feasible and~$(j_a,j_b)$ is a type-inversion on~$\pi'$, we have
$$
c_b \underbrace{\leq}_{(1)} d_{j_b} \underbrace{<}_{(2)} d_{j_a}.
$$
Since~$\pi'_\text{new}$ results from~$\pi'$ by swapping~$j_a$ and~$j_b$, the new completion time of~$j_b$ is~$c_b -t_{j_a} \leq c_b$. Thus, by~$(1)$, job~$j_b$ meets its deadline under~$\pi'_\text{new}$.  Furthermore, the new completion time of~$j_a$ is~$c_b$, and therefore,~$j_a$ meets its deadline by~$(2)$.

Summarizing, swapping~$j_a$ and~$j_b$ inside~$\pi'[I_i]$ results in a feasible schedule that has strictly less type-inversions on~$\pi'[I_i]$ without increasing the makespan. This contradicts the minimality of~$\xi$.

\textbf{Case 2:} $\sigma$ is non-empty\textbf{.} We handle this case by transforming~$\pi'$ into a sequence~$\pi''_\text{new}$ satisfying the constraint of Case~1. Without loss of generality, we assume that no further job of type~$\tau$ is scheduled between~$j_a$ and~$j_b$ on~$\pi'$. Otherwise, we replace either~$j_a$ or~$j_b$ by a corresponding job in between. We define
$$\pi''_\text{new} := \pi'_\text{pre} \circ \sigma \circ (j_a,j_b) \circ \pi'_{\text{suf}}.$$
Intuitively,~$\pi''_\text{new}$ results from~$\pi'$ by interchanging~$j_a$ with the whole block~$\sigma$. Note that we only rearranged jobs inside~$\pi'[I_i]$ and therefore,~$\pi''_\text{new}$ has multi-window distance at most~$k$ from~$\pi$.
Since the jobs on~$\sigma$ do not have type~$\tau$, we have~$x <_{\pi''_\text{new}} y$ if and only if~$x <_{\pi'} y$ for every pair of jobs~$(x,y)$ of the same type. Thus, the number of type-inversions on~$\pi''_\text{new}[I_i]$ and~$\pi'[I_i]$ are the same.

We next consider the makespan of~$\pi''_\text{new}$. To this end, let~$\tau_\text{pre}$ be the type of the last job on~$\pi'_\text{pre}$. If the prefix is empty, we set~$\tau_\text{pre}:=\tau$. Furthermore, let~$\tau_{\sigma}^1$ be the  type of the first job on~$\sigma$, and let~$\tau^2_\sigma$ be the type of the last job on~$\sigma$. Since~$\pi'$ and~$\pi''_\text{new}$ schedule the same jobs, the difference of the makespans can be expressed by the setup times that differ. Formally, this is
\begin{align*}
\Delta &:= C^{\pi''_\text{new}}_{\max} -C^{\pi'}_{\max} \\
&= \left( \ell(\tau_\text{pre},\tau^1_\sigma) + \ell(\tau^2_\sigma,\tau) \right) - \left( \ell(\tau_\text{pre},\tau) + \ell(\tau,\tau_\sigma^1) + \ell (\tau_\sigma^2,\tau) \right).
\end{align*}
Since~$\ell$ satisfies the triangle inequality, we have~$\ell(\tau_\text{pre},\tau) + \ell(\tau,\tau_\sigma^1) \geq \ell(\tau_\text{pre},\tau^1_\sigma)$ and therefore~$\Delta \leq 0$. Consequently, the makespan of~$\pi''_\text{new}$ is not greater than the makespan of~$\pi'$.

It remains to show that each job meets its deadline on~$\pi''_\text{new}$. Obviously, all jobs on~$\pi'_\text{pre}$ meet their deadlines, as they meet their deadlines on~$\pi'$. Next, observe that~$\Delta$ is also the difference between the starting time step of job~$j_b$ on~$\pi''_\text{new}$ and on~$\pi'$. Since~$\Delta \leq 0$, the job~$j_b$ does not start at a later time step as it does on~$\pi'$. Consequently,~$j_b$ and all jobs on~$\pi'_\text{suf}$ meet their deadlines. Since~$j_a$ is scheduled before~$j_b$ and we have~$d_{j_a} > d_{j_b}$, job~$j_a$ also meets its deadline. Finally, since~$\ell(\tau_\text{pre},\tau) + \ell(\tau,\tau_\sigma^1) \geq \ell(\tau_\text{pre},\tau^1_\sigma)$ by the triangle inequality, all jobs in~$\sigma$ do not start at a later time step on~$\pi''_\text{new}$ as they do on~$\pi'$. Summarizing, all jobs meet their deadlines. 

Note that~$\pi''_\text{new}$ satisfies the constraints of Case~1. Thus, by repeating the arguments from the previous case, we obtain a contradiction to the minimality of~$\xi$. Since both cases are contradictory, 
\iflong
Statement~$b)$~holds.
\else
the statement holds for \textsc{Multi-Window LS MM}.
\fi

\iflong
We next consider Statement~$a)$.
\else
We next consider \textsc{Window LS MM}.
\fi
 Let~$\pi'$ be a feasible schedule with window distance at most~$k$ from~$\pi$ such that~$C^{\pi'}_{\max} < C^{\pi}_{\max}$. Furthermore, let~$\xi$ be the number of type-inversions in the rearranged window~$\pi'[a,b]$, such that~$\xi$ is minimal among all such~$\pi'$. Note that in the proof of~$b)$, modifying one single rearranged window~$\pi'[I_i]$ was sufficient to obtain a contradiction to the minimality of~$\xi$ in the case of~$\xi>0$. By repeating the same arguments for the single rearranged window~$\pi'[a,b]$, we conclude~$\xi=0$. 
\iflong
Thus, Statement~$a)$ holds.
\else
Thus, the statement holds for \textsc{Window LS MM}.
\fi
\end{proof}
}
\iflong
\else
\fi

\proofOfObservation

We would like to emphasize that the triangle inequality---which is needed for Proposition~\ref{Prop: EDDS}---is a natural property of setup functions in practice: it appears to be unrealistic that setting up to an intermediate type~$\tau'$ accelerates setting up a machine from a type~$\tau_1$ to another type~$\tau_2$. Even if that was the case, one would declare this intermediate setup as the standard setup from~$\tau_1$ to~$\tau_2$ and consider the updated setup function instead.

\paragraph{Subroutine Inside Windows.} \label{Section: Subroutine}
We now describe how to find the best EDDS inside a window. We later use this algorithm as a subroutine to solve~\textsc{Window LS MM} and \textsc{Multi-Window LS MM}. To formally describe the subroutine, we introduce an auxiliary problem called~\textsc{Internal MM}. The intuition behind this problem is as follows: By Proposition~\ref{Prop: EDDS} we can limit the search space inside one window to EDDSs. Our window contains the jobs of some set~$J_1$. A prefix of the schedule is already known to have some makespan~$\theta$. A suffix of the schedule is known to be some permutation~$\sigma$ of a job set~$J_2$ disjoint from~$J_1$. To keep track of the setup times when using an algorithm for~\textsc{Internal MM} as a subroutine, we keep track of the start type and the end type of the resulting EDDS.

\begin{definition}
Given a job set~$J$ and a setup mapping~$\ell$, a schedule~$\pi$ is called~\emph{$\theta$-feasible} for some integer~$\theta \in \mathds{N}$, if for every~$i$, we have~$C^J_\pi (i)+ \theta \leq d_{\pi(i)}$.
\end{definition}

\taskprob{Internal MM}
{Two disjoint sets~$J_1$ and~$J_2$ of jobs, types~$\tau_1,\tau_2$, a schedule~$\sigma$ of~$J_2$, and some integer~$\theta$.}
{Find an EDDS~$\pi$ of~$J_1$ starting with a job of type~$\tau_1$ and ending with a job of type~$\tau_2$ such that~$\pi \circ \sigma$ is a~$\theta$-feasible schedule for~$J_1 \cup J_2$ and~$\theta + C_{\max}^{\pi \circ \sigma}$ is minimal among all such~$\pi$.}

In the proof of the next proposition, we describe how to solve \textsc{Internal MM} by a dynamic programming algorithm. The algorithm is very  closely related to algorithms for \TTMM~\cite{CK01,GBSB23}. 
\iflong 
We provide it here for sake of completeness. 
\fi
In the following,~$\mathcal{T}$ denotes the set of all types occurring in the set~$J_1$.

\begin{proposition} \label{Prop: Subroutine}
\textsc{Internal MM} can be solved in \iflong{}time
$$\left( \prod_{\tau \in \mathcal{T}} (q_\tau +1) \right) \cdot |I|^{\Oh(1)}.$$\else{}time~$\left( \prod_{\tau \in \mathcal{T}} (q_\tau +1) \right) \cdot |I|^{\Oh(1)}.$\fi{} Herein, \iflong{}$|I|$ is the total input size, and \fi{}$q_\tau$ is the number of jobs with type~$\tau$ in~$J_1$.
\end{proposition}
\newcommand{\proofOfSubroutine}{

\iflong
\begin{proof}
\else
\begin{proof}
\fi
Let~$(J_1,J_2,\tau_1,\tau_2,\sigma,\theta)$ be an instance of \textsc{Internal MM}. Furthermore, let~$\mathcal{T}:= \{1, \dots, t\}$ denote the set of types occurring in~$J_1$. We describe an algorithm based on dynamic programming. Given a type~$\tau$, we let~$\text{EDD}(\tau)$ denote a sequence containing all jobs of type~$\tau$ ordered in non-decreasing manner by their due-dates. Since we aim to find an EDDS~$\pi$, we consider each~$\text{EDD}(\tau)$ as a pre-ordered chain of jobs that keep their relative positions on~$\pi$. The algorithm computes the best possible solution for prefixes of all~$\text{EDD}(\tau)$ in a bottom-up manner, starting with the empty sequence, where all prefixes have length~$0$. To address partial solutions by their prefix-lengths, we use~$t$-dimensional vectors~$\vec{p}$ such that the~$\tau$th entry of~$\vec{p}$ corresponds to the length of the prefix of~$\text{EDD}(\tau)$ that is scheduled in a corresponding partial solution. Furthermore, we let~$\vec{p}-\vec{e}_\tau$ denote the vector that is obtained from~$\vec{p}$ when the~$\tau$th entry is decreased by~$1$.

The entries of our dynamic programming table have the form~$T[\vec{p},\tau]$, where~$\tau \in \mathcal{T}$, and~$\vec{p}$ is a~$t$-dimensional vector with~$p_i \in [0,|\text{EDD}(i)|]$ for every~$i \in \mathcal{T}$. Each entry stores the minimum value of~$C_{\max}^{\pi'}+ \theta$ over all~$\theta$-feasible schedules~$\pi'$ ending with a job of type~$\tau$ and containing exactly the jobs indicated by the prefix-vector~$\vec{p}$. If no such feasible schedule exists, the entry stores~$\infty$.

 We fill the table for increasing values of~$|\vec{p}|:=\sum_{i=1}^t p_i$. For~$|\vec{p}|=0$, that is, for~$\vec{p}=\vec{0}$, we set
\begin{align*}
T[\vec{p},\tau]=
\begin{cases}
\theta &\text{ if }\tau=\tau_1,\\
\infty &\text{ if }\tau\neq\tau_1.
\end{cases} 
\end{align*}
Next, for all~$\vec{p}$ with~$|\vec{p}|>0$ and~$p_\tau=0$, we set~$T[\vec{p},\tau]= \infty$. The recurrence to compute an entry with~$|\vec{p}|>0$ and~$p_\tau>0$~is
$$
T[\vec{p},\tau] := \min_{\substack{\tau' \in \mathcal{T}\\ \text{feas}(\tau',\tau)}} T[\vec{p}-\vec{e}_{\tau},\tau'] + \ell(\tau',\tau) + t_{\text{EDD}(\tau)[p_\tau]}.
$$
Herein,~$\text{feas}(\tau',\tau)$ is either true or false depending on whether the~$p_\tau$th job on~$\text{EDD}(\tau)$ can be scheduled after a job of type~$\tau'$ without violating its deadline. This can be evaluated by checking if~$T[\vec{p}-\vec{e}_{\tau},\tau'] + \ell(\tau',\tau) + t_{\text{EDD}(\tau)[p_\tau]} \leq d_{\text{EDD}(\tau)[p_\tau]}$.

After the table is filled, we consider~$T[\vec{q},\tau_2]$, where~$\vec{q}$ is the~$t$-dimensional vector, where the~$\tau$th entry corresponds to the total number of jobs with type~$\tau$ in~$J_1$. We consider the cases~$J_2 = \emptyset$ and~$J_2 \neq \emptyset$: In the case of~$J_2= \emptyset$, we return the value~$T[\vec{q},\tau_2]$ if it is finite. If~$T[\vec{q},\tau_2]=\infty$, we report that no solution for this instance exists. The corresponding sequence~$\pi$ can be computed via traceback. In the other case, that is~$J_2 \neq \emptyset$, we check whether the schedule~$\sigma$ for~$J_2$ is~$T[\vec{q},\tau_2]+\ell(\tau_2,\tau_{\sigma[1]})$-feasible. If this is the case, we compute the corresponding sequence~$\pi$ via traceback. Otherwise, we report that no solution for this instance exists. Algorithm~\ref{alg:internal-mm} shows pseudo code for the described dynamic programming algorithm.

\begin{algorithm}[t]
	\caption{Pseudo code of the dynamic programming algorithm}
	\label{alg:internal-mm}
	\begin{algorithmic}[1]
		\Input \ Instance $(J_1,J_2,\tau_1,\tau_2,\sigma,\theta)$ of~\textsc{Internal MM}
		\EndInput
		
		\State Extract the set of types~$\mathcal{T}$ and the set of all prefix vectors from the input.
		\State Initialize table~$T$ with entries~$T[\vec{p},\tau]$ for every prefix vector~$\vec{p}$ and every~$\tau \in \mathcal{T}$.
		\State $T[\vec{p},\tau] \gets \infty$ for all entries of~$T$\;
		\State $T[\vec{0},\tau_1] \gets \theta$\;
		
		\ForAll{prefix vectors $\vec{p} \neq \vec{0}$ in increasing order of~$|\vec{p}|$}
			\ForAll{$\tau \in \mathcal{T}$ with $p_\tau > 0$}
				\State $j \gets \text{EDD}(\tau)[p_\tau]$\;
				\ForAll{$\tau' \in \mathcal{T}$}
					\State $x \gets T[\vec{p}-\vec{e}_\tau,\tau']
					+ \ell(\tau',\tau) + t_j$\;
					\If{$x \leq d_j$}
						\State $T[\vec{p},\tau] \gets
						\min\{T[\vec{p},\tau],x\}$\;
					\EndIf
				\EndFor
			\EndFor
		\EndFor
		
		\If{$T[\vec{q},\tau_2] = \infty$}
			\textbf{return} \textsc{No solution}
		\EndIf
		
		\If{$J_2 \neq \emptyset$}
			\If{$\sigma$ is not~$(T[\vec{q},\tau_2]
				+ \ell(\tau_2,\tau_{\sigma[1]}))$-feasible}
				\textbf{return} \textsc{No solution}
			\EndIf
		\EndIf
		
		\State \textbf{return} $\pi$ found via traceback
	\end{algorithmic}
\end{algorithm}

We next consider the running time of the algorithm. The number of table entries is the product of~$|\mathcal{T}|$ and the number of possible prefix vectors~$\vec{p}$. Since the values of the~$\tau$th entry of such prefix vector range from~$0$ to~$q_\tau$, there are at most~$\prod_{\tau \in \mathcal{T}} (q_\tau +1)$ such prefix vectors. Since each entry can be computed in~$|I|^{\Oh(1)}$~time, the algorithm has the stated running time.
\end{proof}
}

\iflong\fi
\proofOfSubroutine

\iflong
The factor~$\prod_{\tau \in \mathcal{T}} (q_\tau +1)$ from the running time relies on the fact that we limit our search space to EDDS. This limitation is done by merging sorted lists for each type. If we were not limiting our search space to EDDS, we can instead `merge' lists of size one. Doing this leads to a running time of~$2^{|J_1|} \cdot |I|^{\Oh(1)}$, which is as good as classic dynamic programming over subsets~\cite{HK62}. Therefore, the algorithm behind~Proposition~\ref{Prop: Subroutine} can be seen as an improvement over a~$2^{|J_1|} \cdot |I|^{\Oh(1)}$-time algorithm by exploiting the solution structure of EDDS.
\fi

Recall that we want to use the algorithm behind Proposition~\ref{Prop: Subroutine} as a subroutine to solve the window local search problems in FPT time for~$k$. To this end, note that the stated running time implies that \textsc{Internal MM} can be solved in~$k^{t}\cdot |I|^{\Oh(1)}$~time, where~$k:=|J_1|$ and~$t$ is the number of different types of jobs occurring in~$J_1$. Note that~$t \leq k$.  This holds due to the fact that the product~$\prod_{\tau \in \mathcal{T}} (q_\tau +1)$ is maximal if all~$q_\tau$ have roughly the same size~$\frac{k}{t}$.

\paragraph{Fixed-Parameter Algorithms.} \label{Section: FPT-Algos}
We now describe how the algorithm behind Proposition~\ref{Prop: Subroutine} together with the observation from Proposition~\ref{Prop: EDDS} can be used to obtain fixed-parameter tractability for~$k$.

\begin{theorem} \label{Theorem: Window-Algo}
\textsc{Window LS MM} can be solved in~$k^{\min(k,t)} \cdot n^{\Oh(1)}$~time, where~$t$ denotes the number of distinct types occurring in the input instance.
\end{theorem}
\newcommand{\proofWinAlgo}{\begin{proof}[Proof of \Cref{Theorem: Window-Algo}.]
The algorithm is straightforward: we iterate over every possible start position of a rearrangement window of length~$k$, and use the algorithm behind Proposition~\ref{Prop: Subroutine} to check if rearranging this window leads to an improvement.

Formally, this is done as follows: For every possible~$i \in [1,n-k]$ and every combination of two types~$\tau'$ and~$\tau''$,\todop{We only need the types of jobs appearing in the window, no? (minor comment)} \todog{In der Implementierung ist es auch so umgesetzt.} we solve the \textsc{Internal MM}-instance
\iflong
$$
(J_1,J_2,\tau',\tau'',\sigma,\theta),
$$
\else 
$(J_1,J_2,\tau',\tau'',\sigma,\theta),$
\fi
where~$J_1$ is the set of jobs on~$\pi[i,i+k-1]$,~$\sigma$ is the (possibly empty) sequence of jobs scheduled after the~$(i+k-1)$th job on~$\pi$, and~$J_2$ is the (possibly empty) set containing the jobs on~$\sigma$. Moreover, we set~$\theta:=0$, if~$i=1$ or~$\theta := C_{\max}^{\pi[1,i-1]} + \ell(\tau_{\pi(i-1)},\tau')$, if~$i>1$. If we found an improvement, we return the improving schedule. Otherwise, we report that no improvement is possible.

The correctness of the algorithm follows by the fact that we consider every combination of~$(i,\tau',\tau'')$, and that it is sufficient to find the best EDDS in a rearranged window due to Proposition~\ref{Prop: EDDS}~$a)$. For the running time, note that each instance of \textsc{Internal MM} can be solved in~$k^{\min(k,t)} \cdot n^{\Oh(1)}$~time. Since we solve at most~$n \cdot t^2$ such instances, the algorithm has the claimed running time.
\end{proof}
}

\iflong
\proofWinAlgo
\else
\begin{proof}[Sketch]
The algorithm is straightforward: we iterate over every possible start position of a rearrangement window of length~$k$, and use the algorithm behind Proposition~\ref{Prop: Subroutine} to check if rearranging this window leads to an improvement.
\end{proof}
\fi

A similar result can be obtained for the multi-window distance by using dynamic programming by exploiting the recursive definition of the decomposition into~$k$-intervals. Intuitively, the recurrence behind the dynamic programming algorithm combines the best ordering of the rightmost window (of size at most~$k$) in the schedule with a solution of the remaining prefix.

\begin{theorem} \label{Theorem: multi-window-Algo}
\textsc{Multi-Window LS MM} can be solved in~$k^{\min(k,t)} \cdot n^{\Oh(1)}$~time, where~$t$ denotes the number of distinct types occurring in the input instance.
\end{theorem}
\newcommand{\proofMultiWin}{
\iflong
\begin{proof}
\else
\begin{proof}
\fi
We show that \textsc{Multi-Window LS MM} is \FPT for\iflong{} the search radius\fi{}~$k$ by providing an algorithm based on dynamic programming. Let~$(J,\ell,\pi,k)$ be an instance of \textsc{Multi-Window LS MM}. 


Before we describe the algorithm, we provide some intuition. When aiming to find the best schedule that has multi-window distance at most~$k$ from~$\pi$ we aim to find the best positions to cut~$\pi$ into windows. These windows must contain at most~$k$ elements. Thus, the last window contains the last~$k' \leq k$ elements on~$\pi$. Consequently, we can find a solution by finding the best cut position~$j$ for the last window and combine it with the best solution for~$\pi[1,j-1]$. To find a solution of the last window, we use the algorithm behind Proposition~\ref{Prop: Subroutine}. We use this recursive solution structure to define a recurrence to fill a dynamic programming~table. 

To formally describe the algorithm, we first introduce some notation. Given some~$\pi[a,b]$, we let~$J(\pi[a,b])$ denote the set of jobs scheduled on~$\pi[a,b]$. Moreover, we let~$\varepsilon$ denote the empty sequence, and we let~$\mathcal{L}(J_1,J_2,\tau_1,\tau_2,\sigma,\theta)$ denote the makespan of a solution of \textsc{Internal MM}.

Our dynamic programming table~$T$ has entries of type~$T[i,\tau]$ with~$i \in [1,n]$ and~$\tau \in \mathcal{T}$. Each entry stores the minimum makespan of a feasible schedule of the jobs on~$\pi[1,i]$ that ends with a job of type~$\tau$ and has multi-window distance at most~$k$ to~$\pi[1,i]$. If no such schedule exists, the entry stores~$\infty$.

The table is filled for increasing values of~$i$. If~$i \leq k$, we~set
$$T[i,\tau] := \min_{\tau' \in \mathcal{T}} \mathcal{L}(J(\pi[1,i]),\emptyset,\tau',\tau,\varepsilon,0).
$$

Note that~$T[i,\tau]$ corresponds to the makespan of an optimal EDDS of~$\pi[1,i]$ ending with type~$\tau$.
For entries with~$i>k$, we set
$$
T[i,\tau]:= \min_{j \in [i-k, i-1]} \min_{\substack{\tau' \in \mathcal{T} \\ \tau'' \in \mathcal{T}}} \mathcal{L}(J(\pi[j+1,i]),\emptyset,\tau'',\tau,\varepsilon,\theta(j,\tau',\tau''))
$$
with~$\theta(j,\tau',\tau''):= T[j,\tau']+\ell(\tau',\tau'')$.

The recurrence described above matches our intuition as follows: We find the best cut position for the last window by iterating over all possible~$j$. We then combine an optimal schedule for this last window~$\pi[j+1,i]$ with a partial solution of the schedule~$\pi[1,j]$ by solving \textsc{Internal MM} for an instance with~$\theta(j,\tau',\tau'')= T[j,\tau']+\ell(\tau',\tau'')$.

After the table is filled, we can compute the~makespan of the solution by evaluating~$\min_{\tau \in \mathcal{T}} T[n,\tau]$. The corresponding schedule can be found via traceback. The correctness of the algorithm follows by the fact that we consider every possible combination of~$(j,\tau',\tau'')$, and that it is sufficient to find the best EDDS in all rearranged windows due to Proposition~\ref{Prop: EDDS}.

We next analyze the running time of the algorithm. The dynamic programming table~$T$ has~$n \cdot t$ entries; recall that~$t$ is the number of distinct types. To compute a single entry, we iterate over each combination of~$j$,~$\tau'$, and~$\tau''$. Consequently, one table entry can be computed with~$k\cdot t^2$ evaluations of the subroutine. Since \textsc{Internal MM} can be solved in~$k^{\min{(k,t)}} n^{\Oh(1)}$~time, the algorithm has the claimed running~time. 
\end{proof}
}

\iflong
\fi

\proofMultiWin

\section{Swap distance and insert distance}\label{sec:4}

In this section, we consider the local search problems~\SWAP and~\INSERT that ask for better schedules with small swap and insert distance, respectively.
To this end, we first formally define these distance measures.

Two distinct schedules~$\pi$ and~$\pi'$ have \emph{swap distance} one, if there are indices~$a\in [1,n]$ and~$b\in [a+1,n]$ such that~$\pi(a) = \pi'(b)$, $\pi(b) = \pi'(a)$, and for each~$i\in [1,n]\setminus \{a,b\}$, $\pi(i) = \pi'(i)$.
Similarly, $\pi$ and~$\pi'$ have \emph{insert distance} one, if there are indices~$a\in [1,n]$ and~$b\in [1,n]\setminus \{a\}$, such that after removing job~$\pi(a)$ from~$\pi$ and inserting it at index~$b$, one obtains~$\pi'$, that is, if~$\pi' = \pi^{-}[1,b-1] \circ (\pi(a)) \circ \pi^{-}[b,n-1]$ for~$\pi^{-} := \pi[1,a-1] \circ \pi[a+1,n]$.


The swap (insert) distance between two schedules $\pi$ and~$\pi'$ equals~$|\mathcal{S}| -1$, where~$\mathcal{S}$ is a shortest sequence of schedules~$(\pi, \dots, \pi')$ that consecutively have swap (insert) distance one.

In the following, we show that for the swap and insert distance, \FPT-algorithms for the search radius~$k$ are unlikely.
\iflong
More precisely, we show that even if each job has processing time one and the given schedule~$\pi$ tardiness zero, \FPT-algorithms for~$k$ plus further natural structural parameters are unlikely.
\else
More precisely, we show that this holds even on very restricted instances.
\fi
On the positive side, we show that a simple brute force algorithm with running time~$n^{2k+1}$ is possible for both distance measures. For our hardness result, we need the standard concept of Hamiltonian cycles, which we first define.

\begin{definition}
For an undirected graph~$G=(V,E)$, a~\emph{Hamiltonian cycle} is a permutation of the vertex set~$C = (v_1, \dots, v_n)$ such that between any two consecutive vertices in~$C$, there is an edge between these vertices in~$G$, where we consider vertex~$v_1$ to follow~$v_n$.
That is, the~\emph{edges of~$C$} defined by $E(C) := \{\{v_i,v_{i+1}\}\mid i\in [1,n-1]\}\cup \{\{v_n,v_1\}\}$ are all contained in~$E$.
For an edge-weight function~$\omega$ and a Hamiltonian cycle~$C$, we denote by~$\omega(C) := \sum_{e\in E(C)}\omega(e)$ the \emph{total weight of all edges of~$C$}.
\end{definition}

\iflong
\begin{theorem}\label{main hardness}
Unless $\FPT = \W1$, neither of \SWAP or \INSERT admits an \FPT-algorithm when parameterized by the search radius~$k$. 
This holds even on instances where simultaneously
\begin{itemize}
\item the initial schedule~$\pi$ has tardiness zero,
\item each job has a deadline of~$\infty$,
\item each job has processing time one,
\item each job has its unique type,
\item the setup mapping assigns only the values~$2,3$, and~$4$, 
\item the setup mapping is symmetrical and fulfills the triangle inequality, and
\item there is an optimal schedule within distance at most~$k$ of the initial schedule~$\pi$.  
\end{itemize}
\todomi{ETH erwähnen}
\todomi{shorter statement in longfalse version}
\end{theorem}
\else
\begin{theorem}\label{main hardness}
Unless $\FPT = \W1$, neither of \SWAP or \INSERT admits an \FPT-algorithm when parameterized by the search radius~$k$, even if each deadline is~$\infty$, each job has processing time one and a unique type, and the setup mapping assign no value larger than~$4$ and is a metric.\todom{okay so?}
All of this holds, even if there is an optimal schedule within distance at most~$k$ of the initial schedule.
\end{theorem}
\fi

\newcommand{\LSTSP}{\textsc{LocalTSP(Swap)}\xspace}

\newcommand\proofOfHardness{
We show the statement for~\SWAP.
The statement for~\INSERT then follows by the fact that two schedules~$\pi$ and~$\pi'$ have insert distance of at most~$2k$ if they have swap distance at most~$k$ and in the \SWAP-instance we construct, the initial solution is~$k$-swap optimal if and only if it is globally optimal.
 
We present a parameterized reduction from~\LSTSP to the decision version of~\SWAP, where the question is: Is there a better schedule with swap distance at most~$k$ from the initial schedule?
\LSTSP is now formally defined as follows.

\prob{\LSTSP}{A graph~$G=(V,E)$ with an edge-weight function~$\omega$, a Hamiltonian cycle~$C$ for~$G$, and an integer~$k$.}{Is there a Hamiltonian cycle~$C'$ for~$G$ with swap distance at most~$k$ from~$C$, such that the total edge-weight of~$C'$ is less than the total edge-weight of~$C$?}

Let~$\widetilde{I}:=(\widetilde{G}=(V,\widetilde{E}),\widetilde{\omega},C,k)$ be an instance of~\LSTSP where each edge has weight either 0 or 1, there is exactly one edge of weight 1, and the initial solution has a globally optimal solution in the~$k$-swap neighborhood.
\LSTSP is~\W1-hard under these restrictions~\cite{GHNS13}.
Note that we can thus assume that the Hamiltonian cycle~$C$ has a total weight of 1, as otherwise, the instance is a trivial no-instance of~\LSTSP.
We describe how to obtain in polynomial time an equivalent instance~$I':=(J,\ell,\pi,k')$ of~\SWAP fulfilling all stated restrictions.

To achieve all of the stated restrictions, we first construct an equivalent instance~$I$ of~\LSTSP having some helpful properties.
Let~$n := |V|$ and let~$G=(V,E)$ be the complete graph on the vertex set~$V$.
We define an edge-weight function~$\omega$ on~$E$ as follows:
for each edge~$\widetilde{e}\in \widetilde{E}$, we set~$\omega(\widetilde{e}) := \widetilde{\omega}(\widetilde{e})$ and for each edge~$e\in E \setminus \widetilde{E}$, we set~$\omega(e) := 2$.
Let~$I:=(G,\omega,C,k)$ be the corresponding instance of~\LSTSP.
Note that each Hamiltonian cycle for~$\widetilde{G}$ is a Hamiltonian cycle for~$G$ of same total weight.
Moreover, each Hamiltonian cycle for~$G$ which uses at least one edge of~$E\setminus \widetilde{E}$ has total weight at least two.
Hence, each globally optimal solution for~$I$ uses only edges of~$\widetilde{E}$, since the Hamiltonian cycle~$C$ has weight~$1$.
This implies that~$I$ and~$\widetilde{I}$ share the same globally optimal solutions and are equivalent instances of~\LSTSP.

Next, we describe how to obtain the equivalent instance~$I'$ of~\SWAP.
Intuitively, we will have for each vertex~$v\in V$ the four jobs~$\vin,\vl,\vr,~\text{and~}\vout$, for which the setup times are defined in such a way that each schedule improving over the initial one needs to keep the four jobs together as the subschedule~$(\vin,\vl,\vr,\vout)$, the subschedule~$(\vout,\vr,\vl,\vin)$, or to start with~$(\vr,\vout)$ ($(\vl,\vin)$) and end with~$(\vin,\vl)$ ($(\vout,\vr)$).
Moreover, the only other job that can follow job~$\vin$ is the job~$\vin[w]$ for some other vertex~$w\in V$, but only if the distance between~$v$ and~$w$ is small enough under~$\widetilde{\omega}$.
In this way, all schedules improving over the initial one correspond to permutations of~$V$ that in turn need to be Hamiltonian cycles of small weight.
Hence, the initial schedule can only be improved if there is a Hamiltonian cycle of smaller weight than the initial one in the local neighborhood.
We now formally define the construction of the instance.
For each vertex~$v\in V$, $J$ contains the four jobs
\iflong
$$\vin,\vl,\vr,~\text{and~}\vout.$$
\else
$\vin,\vl,\vr,~\text{and~}\vout$.
\fi
We call~$\vin$ and~$\vout$ the~\emph{border jobs of~$v$} and~$\vl$ and~$\vr$ the~\emph{center jobs of~$v$}.
As already mentioned, each job has its unique type, unit processing time, and a deadline of~$\infty$.

Next, we describe the setup times between distinct types of jobs. 
Since each job shall have its unique type and the setup matrix shall be symmetrical, we may refer to the setup times between any two distinct jobs of~$J$.
All setup times are from~$\{2,3,4\}$.
We start by describing the setup time between a center job and any other job.
For each vertex~$v\in V$, we define the setup time between~$\vl$ and~$\vin$ as~$2$, the setup time between~$\vl$ and~$\vr$ as~$3$, and the setup time between~$\vl$ and any other job of~$J\setminus \{\vin,\vl,\vr\}$ as~$4$.
Similarly, we define the setup time between~$\vr$ and~$\vout$ as~$2$ and the setup time between~$\vr$ and any other job of~$J\setminus \{\vl,\vr,\vout\}$ as~$4$.
It remains to define the setup times between border jobs.
For each vertex~$v\in V$, we define the setup time between~$\vin$ and~$\vout$ as~$4$.
Finally, for each two distinct vertices~$v$ and~$w$ of~$V$, we define the setup time between~$\vin$ and~$\vout[w]$ as~$2+\omega(\{v,w\})$, the setup time between~$\vin$ and~$\vin[w]$ as~$4$, and the setup time between~$\vout$ and~$\vout[w]$ as~$4$.  
This completes the definitions of all setup times.
Note that the setup matrix fulfills the triangle inequality, since each setup time is at least 2 and at most 4. 

It remains to define the initial schedule~$\pi$ and the search radius~$k'$.
Let~$C'$ be any Hamiltonian cycle for~$G$ with~$C' = (v_1, \dots, v_n)$.
We define a schedule~$\pi_{C'}$ as:
$$\pi_{C'} := (\vr[v_1], \vout[v_1], \vin[v_2], \vl[v_2], \vr[v_2], \vout[v_2], \vin[v_3], \dots, \vout[v_n],\vin[v_1], \vl[v_1]).$$
We now define the initial schedule~$\pi$ as~$\pi:= \pi_C$ and~$k' := 4 k$.
This completes the construction of~$I'$.

The intuitive idea is that each schedule that improves over~$\pi$ is equal to~$\pi_{C'}$ for some Hamiltonian cycle~$C'$ for~$G$ which improves over~$C$ and vice versa.
Moreover, the swap distance between~$\pi = \pi_C$ and~$\pi_{C'}$ is at most four times the swap distance between~$C$ and~$C'$.

\iflong
To show the correctness of the reduction, we make some observations about the structure of schedules that improve over~$\pi$.
To this end, we first analyze the total setup time of~$\pi_{C'}$ for any Hamiltonian cycle~$C'$ of~$G$.
\begin{myclaim}\label{makespan of hamschedules}
Let~$C'$ be a Hamiltonian cycle for~$G$.
The total setup time of~$\pi_{C'}$ is~$9n-3 + \omega (C')$. 
Moreover, $\pi$ and~$\pi_{C'}$ are~$k'$-swap-neighbors if~$C$ and~$C'$ are~$k$-swap-neighbors.
\end{myclaim}
\begin{claimproof}
Let~$C' = (v_1, \dots, v_n)$ and recall that~$\ell$ denotes the setup matrix of~$I'$.
The total setup time of~$\pi_{C'}$ is
\begin{align*}
\sum_{v\in V} & \left(\ell(\vin,\vl) + \ell(\vr,\vout)\right) + \sum_{v\in V \setminus \{v_1\}} \ell(\vl,\vr)\\
+ & \sum_{i=1}^n \ell(\vout[v_{i}],\vin[v_{(i\mod n) + 1}])\\
&= 4 n + 3 \cdot (n-1) + \sum_{i=1}^n (2 + \omega(v_{i},v_{(i\mod n) + 1}))\\
&= 4 n + 3 \cdot (n-1) + 2n + \omega(C') = 9n - 3 + \omega(C').
\end{align*}
Since there are four jobs for each vertex of~$V$, the swap distance between~$\pi$ and~$\pi_{C'}$ is at most four times the swap distance between~$C$ and~$C'$.
\end{claimproof}

Hence, the total setup time of~$\pi$ is~$9n -2$, since~$\omega(C) = 1$.

Next, we show that for each schedule~$\pi'$ that improves over~$\pi$, there is a Hamiltonian cycle~$C'$ of~$G$, such that~$\pi'$ is either~$\pi_{C'}$ or the reverse schedule of~$\pi_{C'}$.
We show this in three steps.
First, we show that each schedule containing a setup time of 4 does not improve over~$\pi$.
Since for each center job~$j$, there are only two jobs from which the setup time to~$j$ is less than 4, this ensures that the jobs corresponding to the same vertex of~$V$ stay together (with at most one exception, where jobs of that vertex may be both at the start and the end of the schedule).
Second, we show that a schedule~$\pi'$ does not improve over~$\pi$ if for each vertex~$v\in V$, $\vr$ directly follows~$\vl$ or vice versa.
Hence, to avoid a setup time of 4, each improving schedule~$\pi'$ starts (ends) with~$\vr$ and ends (starts) with~$\vl$ for some vertex~$v\in V$.
These two facts then imply that~$\pi'$ is either~$\pi_{C'}$ or the reverse schedule of~$\pi_{C'}$.

\begin{myclaim}\label{improve implies light}
Let~$\pi'$ be a schedule of~$J$ such that there are two consecutive jobs~$j_1$ and~$j_2$ on~$\pi'$ such that~$\ell(j_1,j_2) = 4$.
Then, $\pi'$ does not improve over the initial schedule~$\pi$. 
\end{myclaim}
\begin{claimproof}
We show that the total setup time of~$\pi'$ is at least~$9n-2$. 
To this end, we analyze for each job~$j\in J$, the sum~$\alpha(j)$ of setup times surrounding~$j$ on~$\pi'$.
Note that the total setup time of~$\pi'$ is equal to~$\frac{1}{2} \cdot \sum_{j\in J} \alpha(j)$.
To show that~$\pi'$ does not improve over~$\pi$, we bound~$\alpha(\pi) := \sum_{j\in J} \alpha(j)$ from below.
To this end, we bound the~$\alpha$-values for all jobs of~$J$ from below.

By definition, each setup time is at least two.
Hence, for each job~$j\in J$, 
$$\alpha(j) \geq 
\begin{cases}
2&,~j\text{ is the first or the last job of}~\pi'\\
4&,~j\text{ is a border job and not the first or the last job of}~\pi'\\
5&,~j\text{ is a center job and not the first or the last job of}~\pi'
\end{cases}$$ 
by the fact that for each center job~$j'$, there is only one other job having setup time at most~$2$ towards~$j'$.
Moreover, for~$j\in \{j_1,j_2\}$, we obtain 
$$\alpha(j) \geq 
\begin{cases}
4&,~j\text{ is the first or the last job of}~\pi'\\
6&,~\text{otherwise}
\end{cases}.$$ 

Let~$J' := \{\pi(1), \pi(4n), j_1,j_2\}$, let $J_c$ denote the center jobs and let~$J_b$ denote the border jobs.
Then we can bound~$\alpha(\pi)$ from below as follows:
\begin{align*}
\alpha(\pi) &= \sum_{j\in J'} \alpha(j) + \sum_{j\in J_b \setminus J'} \alpha(j) + \sum_{j\in J_c \setminus J'} \alpha(j)\\
&\geq \sum_{j\in J'} \alpha(j) + 4 \cdot |J_b \setminus J'| + 5 \cdot |J_c \setminus J'| \\
&= \sum_{j\in J'} \alpha(j) + 4 \cdot (2n-|J_b \cap J'|) + 5 \cdot (2n-|J_c \cap J'|) \\
&= 18n + \sum_{j\in J'} \alpha(j) - 4 |J_b \cap J'| - 5  |J_c \cap J'| \\
&= 18n + \sum_{j\in J'} \alpha(j) - 4 |J'| - |J_c \cap J'|.
\end{align*}

Recall that at least one of~$j_1$ and~$j_2$ is neither the first nor the last job of~$\pi'$, since~$j_1$ and~$j_2$ are consecutive jobs on~$\pi'$.
Note that this implies that~$\alpha(j_1) + \alpha(j_2) \geq 10$.
Moreover, this further implies that~$|J'|\geq 3$.
If~$|J'| = 3$, then $\sum_{j\in J'} \alpha(j) \geq 11$ which implies that~$\alpha(\pi) \geq  18n + 11 - 15 = 18n - 4$.
Otherwise, if~$|J'| = 4$, we get~$\alpha(j_1) + \alpha(j_2) \geq 12$ and thus $\sum_{j\in J'} \alpha(j) \geq 16$.
Hence, $\alpha(\pi) \geq  18n + 16 - 20 = 18n - 4$.
In both cases, the total setup time of~$\pi'$ is at least~$9n-2$, which implies that~$\pi'$ is not improving over~$\pi$.
\end{claimproof}

In the following, we call a schedule~$\pi'$ of~$J$~\emph{light}, if the setup time between any two consecutive jobs on~$\pi'$ is at most 3.
Recall that for each vertex~$v\in V$, the setup time between~$\vl$ and any job of~$J\setminus \{\vin,\vl,\vr\}$ is 4.
Similarly, the setup time between~$\vr$ and any job of~$J\setminus \{\vl,\vr,\vout\}$ is 4.
Hence, in a light schedule~$\pi'$, for each vertex~$v\in V$, either (a)~$\vl$ directly follows~$\vr$ on~$\pi'$, (b)~$\vr$ directly follows~$\vl$ on~$\pi'$, or (c)~$\pi'$ starts (ends) with~$\vl$ and ends (starts) with~$\vr$.
Next, we show that for each light schedule~$\pi'$ that improves over~$\pi$, there is a vertex~$v\in V$ for which property~(c) holds.

\begin{myclaim}\label{one pair center jobs split}
Let~$\pi'$ be a light schedule of~$J$ such that for each vertex~$v\in V$, $\vl$ directly follows~$\vr$ on~$\pi'$ or vice versa.
Then, $\pi'$ does not improve over the initial schedule~$\pi$. 
\end{myclaim}
\begin{claimproof}
Since~$J$ contains~$4n$ jobs and for each vertex~$v\in V$, $\vl$ directly follows~$\vr$ on~$\pi'$ or vice versa, the total setup time of~$\pi'$ is at least
\iflong $$\sum_{v\in V} \ell(\vl,\vr) + (3n-1)\cdot 2 = 9n-2.$$ \else $\sum_{v\in V} \ell(\vl,\vr) + (3n-1)\cdot 2 = 9n-2$. \fi
Due to~\Cref{makespan of hamschedules}, this is at least the total setup time of~$\pi$ and thus, $\pi'$ does not improve over~$\pi$.
\end{claimproof}

This implies that for each light schedule~$\pi'$ that improves over~$\pi$, there is a vertex~$v\in V$ such that~$\pi'$ starts with~$\vr$ ($\vl$) followed by~$\vout$ ($\vin$) and ends with~$\vl$ ($\vr$) preceded by~$\vin$ ($\vout$).

We are finally ready to show that each schedule that improves over~$\pi$ is either~$\pi_{C'}$ or the reverse of~$\pi_{C'}$ for some Hamiltonian cycle~$C'$ of~$G$.
\begin{myclaim}\label{imp is hamilsched}
Let~$\pi'$ be a schedule for~$J$ that improves over~$\pi$.
Then, there is a Hamiltonian cycle~$C'$ of~$G$ such that~$\pi'$ is either~$\pi_{C'}$ or the reverse of~$\pi_{C'}$.
\end{myclaim}
\begin{claimproof}
Since~$\pi'$ improves over~$\pi$, \Cref{improve implies light} implies that~$\pi'$ is light.
Moreover, \Cref{one pair center jobs split} further implies that there is a vertex~$v_1 \in V$ such that~$\pi'$ starts with~$\vr[v_1]$ ($\vl[v_1]$) followed by~$\vout[v_1]$ ($\vin[v_1]$) and ends with~$\vl[v_1]$ ($\vr[v_1]$) preceded by~$\vin[v_1]$ ($\vout[v_1]$).
Recall that the setup matrix is symmetrical and that all jobs have a deadline of~$\infty$.
Hence, a schedule~$\pi'$ and its reverse schedule both have the same makespan and the same tardiness of zero.
Without loss of generality, we may thus assume that~$\pi'$ starts with~$\vr[v_1]$ followed by~$\vout[v_1]$ and ends with~$\vl[v_1]$ preceded by~$\vin[v_1]$.

Note that since~$\pi'$ is light, for each vertex~$v\in V\setminus \{v_1\}$, $\pi'$ contains either the subschedule~$(\vin,\vl,\vr,\vout)$ or the subschedule~$(\vout,\vr,\vl,\vin)$.
Furthermore, recall that the setup time between $\vout[v_1]$ and any job of~$J\setminus (\{\vr[v_1],\vout[v_1]\}\cup \{\vin\mid v\in V\})$ is at least~$4$.
Hence, for some vertex~$v_2\in V\setminus \{v_1\}$, $\vin[v_2]$ directly follows~$\vout[v_1]$ on~$\pi'$.
By the above, this further implies that the subschedule~$(\vin[v_2],\vl[v_2],\vr[v_2],\vout[v_2])$ directly follows~$\vout[v_1]$ on~$\pi'$.
Applying this argument inductively yields that~$\pi'$ is equal to~$\pi_{C'}$ for some Hamiltonian cycle~$C'$ of~$G$.
\end{claimproof}

Note that this implies that any schedule~$\pi'$ for~$J$ has total setup time at least~$9n-3$ due to~\Cref{makespan of hamschedules}.
Hence, the makespan of~$\pi$ can be improved by at most 1.
Moreover, this implies that each schedule that improves over~$\pi$ is a globally optimal schedule.

Based on the observed properties of improving schedules, we can now prove the correctness.

$(\Rightarrow)$
Let~$C'$ be a Hamiltonian cycle for~$G$ with~$\omega(C') < \omega(C)$ and where~$C$ and~$C'$ are~$k$-swap-neighbors.
Since~$\omega(C) = 1$, this implies that~$\omega(C') = 0$.
Consider the schedule~$\pi_{C'}$ of~$J$.
Due to~\Cref{makespan of hamschedules}, the total setup time of~$\pi_{C'}$ is~$9n-3$ and~$\pi$ and~$\pi_{C'}$ are~$k'$-swap-neighbors.
Hence, $\pi_{C'}$ improves over~$\pi$ and thus, $I'$ is a yes-instance of~\SWAP, where a globally optimal solution is a~$k'$-swap-neighbor of the initial schedule.

$(\Leftarrow)$
Let~$\pi'$ be a schedule of~$J$ that improves over~$\pi$ where~$\pi$ and~$\pi'$ are~$k'$-swap-neighbors.
Due to~\Cref{imp is hamilsched}, we can assume without loss of generality, that~$\pi'$ is equal to~$\pi_{C'}$ for some Hamiltonian cycle~$C'$ of~$G$.
Since~$\pi'$ improves over~$\pi$, \Cref{makespan of hamschedules} implies that~$\omega(C') < \omega(C)$.
Hence, $C$ is not a globally optimal solution for~$I$ which implies that~$C$ is not~$k$-swap-optimal, since the instance~$I$ of~\LSTSP provides the property that~$C$ is a globally optimal solution if and only if~$C$ is~$k$-swap-optimal.
Consequently, $I$ is a yes-instance of~\LSTSP. 
\else

The formal correctness proof is deferred to the appendix.
\fi
}

\iflong
\begin{proof}
\proofOfHardness
\end{proof}
\else
\begin{proof}[Sketch]
\proofOfHardness
\end{proof}
\fi

In other words, even on very restricted instances and for each computable function~$f$, a running time of~$f(k)\cdot |I|^{\Oh(1)}$ for~\SWAP or~\INSERT is unlikely.
Still, a simple brute-force algorithm yields the following polynomial running time for each constant value of the search radius~$k$ for both problems.

\begin{theorem}
\SWAP and~\INSERT can be solved in $n^{2k+1}$~time.
\end{theorem}
\iflong
Essentially this follows by a simple brute-force algorithm that evaluates for all~$k$ consecutive applied swaps (inserts) the solution quality of the obtained schedule.
The latter can be done in~$\Oh(n)$ time and enumerating all~$k$ consecutive applied swaps can be done in $n^{2k}$~time.
\fi

\section{Derived hill climbing algorithms} \label{Section: Hill-Climbing}
Based on the theoretical results of the previous sections, we derived four hill climbing local search algorithms: Win, Win+Swap, MW, MW+Swap.
To motivate the algorithms we want to refer to the state-of-the-art hill-climbing algorithm PILS1 proposed by Rego et al.~\cite{RSA12}.
Roughly speaking, PILS1 improves schedules by using swaps and inserts to perform local search steps combined with a perturbation of the schedules when a local optimum was reached. The perturbation consists of randomly reversing the order of~$k$ consecutive jobs on the schedule. If no improvement was found, the value of~$k$ increases. 

In contrast to the use of random perturbations, the idea behind our hill-climbing strategies is to use parameterized local search as a deterministic subroutine to reorder~$k$ consecutive jobs on a schedule. Analogously to PILS1, the value of~$k$ increases if no improvement was found. We also combine this idea with swaps as simple local search steps.

\paragraph{The Win and MW Algorithms.} The general idea of these hill climbing algorithms is to start with some initial solution~$\pi$ and~$k=4$.
The algorithm then tries to replace the current solution by a better one with window distance or multi-window distance at most~$k$.
To evaluate whether there is a better solution with distance at most~$k$, we use the algorithm behind Theorems~\ref{Theorem: Window-Algo} and~\ref{Theorem: multi-window-Algo}. 
If no such better solution exists, the algorithm is stuck in a locally optimal solution.
Aiming to escape this locally optimal solution, the value of~$k$ is incremented, which yields a larger neighborhood to consider.
Otherwise, if there is a better schedule~$\pi'$ with distance at most~$k$, we replace~$\pi$ by~$\pi'$ and set~$k$ back to~$4$.
We always set~$k$ back to~$4$ after finding any better solution, because searching for a better solution with distance at most~$k$ can be performed faster the smaller~$k$ is.
We chose the reset value of~$k$ to be~$4$ instead of~$2$ because preliminary experiments showed that for~$k\in\{2,3\}$, improvements happened rarely.

\paragraph{The Win+Swap and MW+Swap Algorithms.} The algorithms Win+Swap and MW+Swap extend Win and MW by additionally searching for better solutions with swap distance~1.
We limit the algorithms to only search for swap distance~1, since the hardness result of~\Cref{main hardness} implies that considering larger swap distances tends to become inefficient.
This was also validated by preliminary experiments when considering also better solutions with swap distance~2 and~3.
Further, preliminary experiments showed that considering the swap distance provided much better solutions than considering the insert distance.
As a consequence, we do not include combinations of insert distance with any of window distance or multi-window distance in our evaluation.

 The pseudo code for Win+Swap is depicted in~\Cref{alg:hillclimbing win plus swap}. 
 Lines~\ref{start swap code}--\ref{end swap code} of~\Cref{alg:hillclimbing win plus swap} check for a better schedule in the~$1$-swap-neighborhood.
\todog{Diese Erklärungen brauchen wir doch jetzt nicht mehr, oder?} 
Omitting these lines yields our second algorithm: Win.
By replacing in Line~\ref{if win opt} and Line~\ref{replace win neighbor} the window distance by the multi-window distance yields MW+Swap.

\begin{algorithm}[t]
	\caption{Pseudo code of Win+Swap}
	\label{alg:hillclimbing win plus swap}
	\begin{algorithmic}[1]
		\Input \ Instance $(J,\ell)$ of~\TTMM and an initial schedule~$\pi$
		\EndInput
		\State $k \gets 4$\;
		\While{time is not exceeded}
		\If {$\pi$ is not~$1$-swap optimal}\label{start swap code}
		\State $\pi\gets$ better schedule with swap distance one
		\State $k\gets 4$
		\State \textbf{continue}\;
		\EndIf\label{end swap code}
		\If {$\pi$ is~$k$-window optimal}\label{if win opt}
		\State $k \gets k+1$\;
		\Else
		\State $\pi\gets$ better schedule with window distance~$\leq k$ \label{replace win neighbor}
		\State $k \gets 4$
		\EndIf
		\EndWhile
		\Return $\pi$
	\end{algorithmic}
\end{algorithm}

\paragraph{Tardiness.}
Note that providing a feasible starting solution with tardiness zero can not be done in polynomial time, unless~$\Ptime = \NP$.
Hence, in our experimental evaluation, we also allow non-feasible schedules.
Then, our goal is to minimize the total tardiness and the makespan, where minimizing the total tardiness is the primary objective. 
That is, a schedule~$\pi'$ is~\emph{better} than a schedule~$\pi$, if~a) the total tardiness of~$\pi'$ is smaller than the total tardiness of~$\pi$ or~b)~$\pi$ and~$\pi'$ have the same total tardiness and the makespan of~$\pi'$ is smaller than the makespan of~$\pi$.
Note that under this modified objective function, if the current solution~$\pi$ has total tardiness zero, the algorithms for all distance measures are still correct.
If the total tardiness of~$\pi$ is non-zero, the previously described algorithms for finding better solution with window distance or multi-window distance are not necessarily correct anymore, that is, the described algorithms output that there is no better solution with distance at most~$k$ even though there might be a better solution with distance at most~$k$, where the corresponding change is not EDDS.
In our evaluation, we show that even though some local improvements might not be found, the approaches still perform very well when additionally aiming to minimize the total tardiness also for non-feasible solutions. 

\paragraph{Considered Starting Solutions.} We consider three starting solutions for our algorithms.
All these starting solutions are computed greedily and are based on the idea of EDDS.
The first starting solution (DD) sorts all jobs in non-decreasing manner by their deadline.
The other two solutions (SM and TM) group the jobs by their type. More precisely, jobs of the same type are consecutive and ordered non-decreasingly by deadlines.
In SM we take an ordering of the type blocks that minimize the total setup times among all such orderings.
In TM we take an ordering of the type blocks that minimizes the total tardiness among all such orderings.
\iflong
The latter two starting solution can be computed in $t!\cdot n + n\cdot \log(n)$~time, where~$t$ denotes the number of types and~$n$ denotes the number of jobs.
\fi 
Since we assume that~$t$ is a very small constant (in our experiments, $t=8$) these starting solutions can be computed efficiently.

\section{Preliminary Experiments}
We performed experiments to evaluate the hill climbing algorithms Win, Win+Swap, MW, and MW+Swap that were described in Section~\ref{Section: Hill-Climbing}. To put the experimental results into context, we also evaluated our implementations of baseline algorithms. These algorithms are the state-of-the-art hill-climbing algorithm PILS1~\cite{RSA12} and two genetic algorithms, as genetic algorithms form an important class of scheduling heuristics in practice.
\iflong
We first describe the details of the baseline algorithms. Afterwards, we discuss the experimental results.
\else 
The description of these algorithms is deferred to the appendix.
\fi 

\iflong
\subsection{Baseline Algorithms}\label{subsec baseline}

\paragraph{The state-of-the-art algorithm PILS1.}
\fi

\newcommand{\descriptionPILS}{
We compare our algorithms with the state-of-the-art algorithm ``PILS1'' proposed by Rego et al.~\cite{RSA12}.

PILS1 takes a set~$\ND$ of pairwise non-dominating schedules, where a schedule~$\pi$ is dominated by another schedule~$\pi'$ if both the tardiness of~$\pi$ is worse than the tardiness of~$\pi'$ and the makespan of~$\pi$ is worse than the makespan of~$\pi'$. Recall that PILS1 improves schedules in~$\ND$ by using swaps and inserts to perform local search steps combined with a perturbation of the schedules in~$\ND$ when a local optimum was reached.

More precisely, until a given time limit is reached, the algorithm takes a schedule~$\pi$ from~$\ND$ uniformly at random and tries to find a schedule~$\pi'$ which is not dominated by any schedule of~$\ND$ such that~$\pi$ and~$\pi'$ have swap distance~1 or insert distance~1 via bruteforce.
If such a solution~$\pi'$ is found, $\pi'$ is added to~$\ND$ and all schedules of~$\ND$ that are dominated by~$\pi'$ are removed from~$\ND$.

Otherwise, if no such schedule~$\pi'$ is found, the algorithm again takes a schedule~$\pi$ of~$\ND$ uniformly at random and performs a perturbation step.
In this perturbation step,  a subschedule of $p$ consecutive jobs of~$\pi$ chosen uniformly at random is reversed within~$\pi$.
The resulting schedule~$\pi'$ is then the next candidate to check for possible neighbors that are not dominated by any  schedule of~$\ND$ with respect to a single swap or insert operation.
Initially, the value of~$p$ is 4.
This value will be restored whenever a new schedule is added to~$\ND$.
If an iteration of the loop does not add a new schedule to~$\ND$, the value of~$p$ is instead incremented by 1.
Finally, when the time limit is exceeded, the algorithm outputs (for our purpose) the schedule of~$\ND$ of lowest tardiness.

Note that the hill climbing strategies Win+Swap and MW+Swap from Section~\ref{Section: Hill-Climbing} are related to PILS1: 
We escape poor local optima by reordering consecutive jobs on the schedule. 
The value of~$p$ in the perturbation step acts similarly to our local search radius for window and multi-window operations, as it is increased when no improvement was found. 
In this sense, Win+Swap and MW+Swap can be seen as an adaption of PILS1 by replacing random perturbations by deterministic steps reordering consecutive jobs on the schedule.}

\iflong
\descriptionPILS
\fi

\newcommand{\explainGenetic}{
\paragraph{Genetic Algorithms.}
For a detailed introduction into genetic algorithm, we refer to the standard monograph~\cite{G89}.
Additionally to PILS1, we consider two genetic algorithms as a baseline. The algorithms maintain a population of size~$p$ 
and consist of standard operators Encode ($E_{n}$), Decode ($D_{e}$), Crossover ($C_{o}$), Mutation ($M_{u}$), 
Evaluate ($E_{v}$), and Selection ($S_{e}$). 
We propose a basic genetic algorithm (GAD)
and a modified genetic algorithm (MGA) that exploits the solution structure from Proposition~\ref{Prop: EDDS}.
To evaluate the quality of the results,
we define the \emph{fitness function} 
as~$\Phi(\pi):=  C_{\max} + \mu T$, where~$T$ is the total tardiness of $\pi$, with~$\mu := n \cdot (\max_{j \in J} t_j + \max_{i,j \in J} \ell (\tau_i,\tau_j)) + 1$ for the job set~$J$. Note that $\Phi(\pi)$
essentially models a lexicographic optimization, where the primary goal is to minimize
the total tardiness and the secondary goal is to minimize the makespan.

First, consider GAD. For the encoding, we fix an arbitrary schedule~$\pi$ and 
use the indices to identify the corresponding jobs by integer numbers. 
We then generate the initial population by creating~$p$ permutations~$\pi^{\prime}$ independently at random. 
All of these permutations are EDDS. The operator $E_{n}$
is used to encode all permutations by storing sequences of the unique numbers. 
Based on preliminary experiments,
we chose a two point crossover method for~$C_{o}$ 
and a single swap method for~$M_{u}$.
To evaluate, all sequences are decoded and scored with~$\Phi(\pi)$. In the last step, 
the $p-1$ best schedules with respect to~$\Phi(\pi)$ are selected in the operator $S_{e}$ and the best 
schedule of the last iteration is copied.

Second, consider MGA. We derived the idea for the MGA based on Proposition~\ref{Prop: EDDS}, that is, 
we limit the solution space to EDDS. It differs from the GAD in the operators $E_{n}$, $C_{o}$, $M_{u}$ and $D_{e}$. 
In case of the MGA, we generate the starting population in a similar way but we encode it by storing the type~$\tau_{j}$ of each job instead of the unique number.
We adapt~$C_{o}$ by using a type 
based crossover. In contrast to randomly choosing a position, 
we randomly select a set containing at most~$k$ types.
Then, all types that occur in the set are copied from the second parent into a new sequence 
and the empty spaces are filled with all other types from the first parent 
maintaining their relative order. For the operator~$M_{u}$, we randomly 
choose two indices and shuffle the subsequence lying in between. 
In addition, the sequences are decoded by mapping the~$\tau_{j}$ and the relative position to the individual jobs grouped by~$\tau_{j}$ and sorted in non-decreasing manner by
deadline~$d_{j}$. With this decoding, all solutions are EDDS.
In the final step, these solutions are evaluated and selected in the same way as for the~GAD.

We implemented both algorithms using the Python library \texttt{PyGad}. Based on preliminary experiments, we chose the parameters of both algorithms as follows: the size of the starting population is~$p := 100$, the crossover probability is~$c_{w} := 0.9$, the mutation probability is~$m_{w} := 0.01$ and the selection rate is $s_{r} := 0.2$.
}

\iflong 
\explainGenetic
\fi

\newcommand{\tablAll}{
\begin{tabular}{m{1.2cm}m{.6cm}|rrrrrr||rrrrrr}
\hline
  &   & 50 & 100 & 150 & 200 & 250 & 300& 50 & 100 & 150 & 200 & 250 & 300\\\hline\hline
Win & DD  & 735 & 1115 &  \cellcolor{black!10}1925 &  \cellcolor{black!10}4245 & 3835 & 5040& 889 & 8434 & 15213 & 586 & 15816 & 54422\\
 & SM  & 715 & 1155 & 2965 & \dag &  \cellcolor{black!10}1450 &  \cellcolor{black!10}750& 889 & 8420 & 15357 & 913 & 16781 & 54634\\
 & TM  & 725 & 1410 & 2260 & \dag & 1705 & 2630& 890 & 8429 & 15178 & 773 & 16376 & 55052\\\hline
Win{\tiny +}Sw & DD  & 755 &  \cellcolor{black!10}1075 & 3195 & 4370 & 3365 & 4055& 730 & 7296 &  \cellcolor{black!10}10774 & 505 &  \cellcolor{black!10}\emph{13772} &  \cellcolor{black!10}\emph{44903}\\
 & SM  & 715 & 1320 & 3010 & \dag & 2080 & 2015& 731 &  \cellcolor{black!10}7251 & 10845 &  \cellcolor{black!10}475 & \emph{18784} & \emph{49399}\\
 & TM  & 710 & 1335 & 2660 & \dag & 2125 & 1775& 735 & 7276 & 10806 & 491 & \emph{17365} & \emph{47011}\\\hline
MW & DD  & 1040 & 2605 & 4660 & 6430 & 8160 & 9955& 953 & 8572 & 15682 & 649 & 16610 & 56935\\
 & SM  & 795 & 1645 & \dag & \dag & 2285 & \dag& 1000 & 8798 & 16315 & 1162 & 22320 & 59634\\
 & TM  & 915 & 2060 & \dag & \dag & 3435 & 4325& 910 & 8651 & 15733 & 897 & 19063 & 59019\\\hline
MW{\tiny +}Sw & DD  & 1000 & 1675 & 3955 & 5480 & 4625 & 7265& 736 & 7323 & 10869 & 527 &  \cellcolor{black!10}\emph{13772} &  \cellcolor{black!10}\emph{44903}\\
 & SM  & 970 & 1635 & 3140 & \dag & 2905 & 2075& 732 & 7308 & 10919 & 478 & \emph{18784} & \emph{49399}\\
 & TM  & 710 & 1840 & 3440 & \dag & 2270 & 2145& 732 & 7316 & 10865 & 497 & \emph{17365} & \emph{47011}\\\hline
PILS1 &  &  \cellcolor{black!10}705 & 1210 & 7455 & 19625 & 10175 & 9630&  \cellcolor{black!10}727 & 7675 & 12332 & 631 & 16316 & 48770\\\hline
GAD &  & 3695 & 12970 & 18810 & \dag & 32850 & 46130& 810 & 8039 & 12823 & 1247 & 18512 & 48864\\\hline
MGA &  & 1125 & 2810 & 5870 & 9540 & 10740 & 15250& 904 & 8745 & 15975 & 699 & 17185 & 58540
\end{tabular} }

\iflong
\begin{table}[!t]
\caption{Experimental results. Subtable~$1)$ displays the total setup time of the best feasible schedules that were found within the time limit. In cells marked with~\dag, no feasible schedule was found within the time limit. Subtable~$2)$ provides an overview on the tardiness of the best schedule found within the time limit. For sake of readability, the table displays~$\lfloor T \cdot 10^{-3} \rfloor$ for every computed tardiness~$T$. The best found solution(s) of each instance are marked with grey cell color. Italic numbers indicate that the stated total tardiness was obtained only by 1-swaps and no other local neighborhood was considered within the time limit.} \label{Table: Experiments}

\begin{center}
\iflong
$1)$ dataset: feasible schedule possible \\
\tableeqLines 

\medskip

$2)$ dataset: feasible schedule impossible\\
\tablegeLines
\else
~~~~~~~~~~~~~~~~~~~~~$1)$ feasible schedule possible~~~~~~~~~~~~~~~$2)$  feasible schedule impossible~~~~~~~\\
\scalebox{0.9}{
\tablAll
}
\fi
\end{center}
\end{table}
\fi

\iflong
\subsection{Experiments}
\fi

We implemented the above mentioned algorithms in Python and evaluated their performance.
The experiments were run with Python 3.10.12 on a single thread of an Intel(R) Xeon(R) E5-2697 v3 CPU with 2.6 GHz and 32 GB RAM.
 
\paragraph{The considered data and setup mapping.}
We perform experiments to evaluate our approaches for the real-world use case from the company~\COMP. The products produced by~\COMP are available in eight different color/material combinations, all of which are manufactured on the same injection molding machine. 
For our experiments, we use an~$8 \times 8$ setup matrix containing the machine's setup times of a real-world injection molding machine in minutes as integers. 

Since \COMP{}s web shop has not yet launched, the total number of real-world orders is relatively small at this time. 
To evaluate our algorithms on large instances, we adapted well-established benchmark datasets for single-machine scheduling with deadlines established by Tanaka et al.~\cite{TFA09}
. 
More precisely, for each job consisting of a processing time and a deadline, we chose a product type from the interval~$[1,8]$ independently at random from an equal distribution. 
Furthermore, we multiplied the processing time and the deadline of each job by a constant of~$50$, since otherwise the real-world setup times from \COMP always dominate the solution's makespan by far, which is not realistic in our use case.
\newcommand{\dataSet}{
\begin{center}
\footnotesize
\begin{tabular}{r|cccccc} Dataset & 50 & 100& 150 & 200& 250 & 300\\ \hline 
$T = 0$ & wt050\_078 & wt100\_105 & wt150\_108 & wt200\_106 & wt250\_088 & wt300\_125 \\
$T > 0$ & wt050\_012 & wt100\_073 & wt150\_017  &  wt200\_005 & wt250\_076 & wt300\_052 
\end{tabular}
\end{center}
}
\iflong
The concrete tested datasets are the following:
\dataSet
\else
In the appendix, we provide an overview of the concrete tested data sets.
\fi

We consider two sets of data. 
One dataset, where a feasible schedule exists, and one dataset, where no feasible schedule exists. 
Recall that, in the second case, we then aim to minimize the total tardiness. 
In both datasets, we have instances with~$n\in \{50, 100, 150, 200, 250, 300\}$ jobs.


\paragraph{Experimental Results.}
For all three starting solutions, we evaluated our four hill climbing strategies \iflong{}(Win, Win+Sw, MW, MW+Sw) \fi{}on~$1)$ a dataset of instances where feasible schedules exist, and~$2)$ a dataset of instances where no feasible schedule exists. Each instance ran within a time limit of 20 minutes.
Furthermore, as a baseline, we ran experiments with the two genetic algorithms GAD and MGA and with the PILS1 algorithm. 
Since GAD, MGA, and PILS1 are non-deterministic randomized algorithms, we performed 20 independent repetitions with a time limit of 20 minutes each and kept the best found results.

We summarize our main experimental findings as follows: 
First, all hill climbing algorithms provide good results for instances where feasible schedules are possible. 
The algorithms were mostly able to provide good results even if the starting solution was not feasible. 
Overall, the window algorithm appears to be well-suited to handle these instances. 
Second, on instances where no feasible schedules exist, local search via single swaps often has the biggest influence on the solution quality and provides good results: 
Especially for the large instances with~$n \in \{250,300\}$, all improvements were reached by swaps and no parameterized local search step has been applied within the time limit of 20 minutes. 
On the smaller instances, the VNS algorithms Win+Sw and MW+Sw help the local search via swaps to escape bad swap-optimal solutions and provide good results for this case.  Overall, Win+Sw appears to be well-suited to handle these instances.
Third, the hill climbing algorithms outperform the baseline genetic algorithms in the solution quality. On the small instances~($n = 50$), PILS1 performed slightly better than the hill-climbing algorithms, while on larger instances~($n\geq 100$), the parameterized local search algorithms mostly provide better results than PILS1. This might indicate that parameterized local search is a reasonable alternative to the perturbation of schedules to escape poor local swap-optimal solutions.


\section{Conclusion}
\iflong
Motivated by a real-world use case, we initiated the study of parameterized local search heuristics for single-machine scheduling from a theoretical and practical point of view. 
%
Our experimental evaluations indicate that parameterized local search is a promising technique in the design of heuristics for the $1 \mid \text{ST}_\text{sd,f}, \overline{d_j} \mid C_{\max}$ scheduling problem. In comparison to the PILS1 algorithm~\cite{RSA12}, our experiments indicate that parameterized local search via windows of consecutive jobs can be a good alternative to random perturbation in a VNS. Besides the positive results, we also theoretically outlined the limits of parameterized local search via insert and swap operations.
\fi

There are many ways to extend our results in future work. Memetic algorithms~\cite{M89} combine local search strategies with genetic algorithms. 
How do memetic algorithms work when the local search steps are performed with~$k>1$? 
For which~$k$ do they deliver a good trade-off between solution quality and running time? 
Considering the parameterized complexity, we may ask whether \SWAP and \INSERT are FPT for the sum of~$k$ and the number~$t$ of distinct types in the instance. Note that the number of types in the instance constructed in the proof of Theorem~\ref{main hardness} is unbounded in the parameter. Thus, FPT for~$k+t$ is still possible. This parameterization would be particularly interesting, since~$t$ is often relatively small in industrial applications (recall that we have~$t=8$ in our \COMP use case). 
Furthermore, it might be interesting to see whether the results in this work can be extended to other target functions than the makespan: is it possible to obtain similar results if one aims to minimize the (weighted) total tardiness or the maximum tardiness?
\iflong
\fi
\bibliographystyle{plain} 
\bibliography{MWLS}

\iflong\else
\newpage
\appendix
\longtrue
\section{Appendix}

\subsection{Missing Proofs of Section~\ref{sec:3}}
\proofOfObservation



\proofWinAlgo

\subsection{Missing Proofs of Section~\ref{sec:4}}
\begin{proof}[Complete proof of~\Cref{main hardness}.]
\proofOfHardness
\lncsqed
\end{proof}

\subsection{Details about PILS1}
\descriptionPILS

\subsection{Details about Genetic Algorithms}
\explainGenetic

\begin{table}[!t]
\caption{Experimental results. Subtable~$1)$ displays the total setup time of the best feasible schedules that were found within the time limit. In cells marked with~\dag, no feasible schedule was found within the time limit. Subtable~$2)$ provides an overview on the tardiness of the best schedule found within the time limit. For sake of readability, the table displays~$\lfloor T \cdot 10^{-3} \rfloor$ for every computed tardiness~$T$. The best found solution(s) of each instance are marked with grey cell color. Italic numbers indicate that the stated total tardiness was obtained only by 1-swaps and no other local neighborhood was considered within the time limit.} \label{Table: Experiments}

\begin{center}
\iflong
$1)$ dataset: feasible schedule possible \\
\tableeqLines 

\medskip

$2)$ dataset: feasible schedule impossible\\
\tablegeLines
\else
~~~~~~~~~~~~~~~~~~~~~$1)$ feasible schedule possible~~~~~~~~~~~~~~~$2)$  feasible schedule impossible~~~~~~~\\
\scalebox{0.9}{
\tablAll
}
\fi
\end{center}
\end{table}

\begin{table}[h]
\caption{The concrete tested data sets provided by Tanaka et al.~\cite{TFA09}.}
\dataSet
\end{table}

\fi

\end{document}